\documentclass[twocolumn]{aastex701}

\usepackage{empheq}
\usepackage{bm}
\usepackage{booktabs}
\usepackage{multirow}
\let\tablenum\relax
\usepackage{siunitx}
\DeclareSIUnit{\kms}{\kilo\meter\per\second}
\usepackage{color,soul}
\usepackage{cleveref}
\crefname{figure}{Figure}{Figures}
\crefname{equation}{Equation}{Equations}
\crefname{table}{Table}{Tables}
\crefname{section}{Section}{Sections}
\crefname{subsection}{Subsection}{Subsections}
\crefformat{equation}{Equation~#2#1#3}
\crefformat{section}{Section~#2#1#3}
\crefformat{subsection}{Subsection~#2#1#3}

\newcommand{\lt}{\left}
\newcommand{\rt}{\right}
\newcommand{\e}{\mathrm{e}}
\newcommand{\gp}{\gamma_{+}}
\newcommand{\gpp}{$\gamma_{+}$ }
\newcommand{\gx}{\gamma_{\times}}
\newcommand{\gxx}{$\gamma_{\times}$ }
\newcommand{\ts}{\textsuperscript}

\makeatletter

\newcommand{\AAS@latexml@patches}{%
  \let\AAS@orig@author\author
  \renewcommand{\author}[2][]{\AAS@orig@author{##2}}%

  \let\AAS@orig@email\email
  \renewcommand{\email}[2][]{\AAS@orig@email{##2}}%

  \@ifundefined{correspondingauthor}{%
    \newcommand{\correspondingauthor}[1]{}%
  }{%
    \renewcommand{\correspondingauthor}[1]{}%
  }%
  \@ifundefined{uat}{%
    \newcommand{\uat}[2]{\href{https://astrothesaurus.org/uat/##2}{##1\ (##2)}}%
  }{%
    \renewcommand{\uat}[2]{\href{https://astrothesaurus.org/uat/##2}{##1\ (##2)}}%
  }%
}

\@ifundefined{LaTeXML}{}{%
    \AAS@latexml@patches
  }%

\makeatother

\newcommand{\rev}[1]{{\color{black}#1}}

\begin{document}

\title{Improving Precision in Kinematic Weak Lensing with MIRoRS: Model-Independent Restoration of Reflection Symmetries}

\author[0000-0001-8559-4538]{Christopher Hopp}
\affiliation{Department of Physics \& Astronomy, University of California, Davis, CA 95616}
\email[show]{chopp@ucdavis.edu}
\correspondingauthor{Christopher Hopp}

\author[0000-0002-0813-5888]{David Wittman}
\affiliation{Department of Physics \& Astronomy, University of California, Davis, CA 95616}
\email{dmwittman@ucdavis.edu}

\begin{abstract}
We present a novel, model-independent technique for fitting the cross-component of weak lensing shear, $\gamma_\times$, along a line of sight by combining kinematic and photometric measurements of a single lensed galaxy. Rather than relying on parametric models, we fit for the shear parameter that best transforms the velocity field to restore its underlying symmetries, while also incorporating photometric data for the change in position angle due to shear.  We first validate our technique with idealized mock data, exploring the method's response to variations in shear, position angle, inclination, and noise. On this idealized mock data, our combined kinematic and photometric model demonstrates superior performance compared to traditional parametric or kinematic-only approaches. \rev{We also explore the effects of asymmetric warps and show that rotation direction can impart a small bias on the fit of $\gamma_\times$.} Subsequently, we apply our method to a dataset of 358 halos from the Illustris TNG simulations, achieving a notable reduction in the uncertainty of $\gamma_\times$ to 0.039, marking a substantial improvement over previous analysis of the dataset with a parametric model. Finally, we introduce an outlier rejection method based on Moran’s $I$ test for spatial autocorrelation. Identifying and filtering out halos with spatially correlated residuals reduces the overall uncertainty to 0.028. Our results underscore the efficacy of combining kinematic and photometric data for weak lensing studies, providing a more precise and targeted measurement of shear along an individual line of sight. 
\end{abstract}


\section{Introduction}\label{sec:Introduction}
Weak gravitational lensing provides a robust method for probing the mass distribution of the Universe (see \citet{bartelmann_2001} for a comprehensive review and \citet{bartelmann_2017} for a more contemporary treatise). As the light from an extended background source travels through the potential well of a foreground lens, it is distorted such that the image observed carries information about the underlying mass distribution of the lens. This distortion manifests in two primary effects: convergence and shear. Convergence, typically denoted $\kappa$, produces an isotropic scaling of the image, maintaining the shape while scaling the size. Shear, denoted $\gamma$, distorts the shape of the image, elongating along one axis and compressing in an orthogonal direction. \rev{This shear can be described by two components: $\gp$ elongates and compresses along the coordinate axes, while the cross-component $\gx$ operates along the line $y=x$.} Determining the intrinsic size of a source is typically difficult; thus, the primary information available is in the ellipticity and orientation of the image.

In the weak lensing regime, these effects are small, typically with ellipticity distortions ${<}1\%$ \citep{mckay2001galaxy}. Meanwhile, the intrinsically wide distribution of unlensed galaxy orientation and ellipticity, or shape noise, overwhelms the signal of a single source. In traditional weak lensing, shape noise is overcome by using a statistical ensemble of sources to search for this systematic distortion. While the uncertainty in the shape of any single source is large, it is presumed that there is no preferential orientation of an unlensed source ensemble. Averaging over a large number of sources, of the order of $\sim 10^4$ in cluster-galaxy lensing \citep{umetsu2020cluster,umetsu2014clash} and up to $\sim 10^7$ in cosmic-shear studies \citep{troxel2018dark}, is required to achieve a desirable signal-to-noise ratio.

With no a priori knowledge of the shape and orientation of an unlensed galaxy, traditional weak lensing will always be plagued by shape noise. However, some source galaxies possess sufficient symmetries in their velocity field such that the light observed is not completely devoid of information on the unlensed condition of the galaxy. From the equivalence principle, it is known that lensing effects are independent of photon wavelength. Thus, the velocity field of a galaxy will also show the effects of shearing. Under idealized conditions, an unlensed disk galaxy has perpendicular kinematic axes and displays symmetry in its velocity field along its major axis and antisymmetry along its minor axis. These symmetries are broken under shear, and it was first recognized by \citet{blain_2002} that this deviation produces an observable that can be used to infer the shear parameter from a single galaxy.

This work was extended by \citet{morales_2006}, and significant progress has since been made in the subfield of Kinematic Lensing\footnote{\rev{We adopt the term Kinematic Lensing broadly, rather than referring to the exact methods introduced in \citet{huff2013cosmic}. Conceptually similar approaches have gone by many different names including direct shear mapping \citep{de_burgh-day_2015}, precision weak lensing \citep{gurri_2020}, and velocity field lensing \citep{donet_2023}.}} (henceforth KL). \rev{Following these early developments, KL methods have evolved in two primary directions shaped by their scientific goals. One avenue, pursued largely in the context of cosmic-shear surveys, aims to extract ensemble-average shear using minimal spectroscopic information \citep{huff2013cosmic, xu2023kinematic, pranjal_2022, huang2025cosmic}. A second approach focuses on using spatially resolved velocity fields of individual galaxies to obtain precise per-source shear constraints, enabling detailed studies of localized shear \citep{de_burgh-day_2015, gurri_2020, digiorgio_novel_2021, donet_2023}.

In both applications, the most common approach involves fitting a parametric model of a sheared velocity field to observations.} This approach was most recently explored by \citet[][hereafter DW23]{donet_2023}, where an uncertainty in the cross-component of shear\footnote{\rev{In parametric modeling of an individual galaxy, the coordinate system is most often chosen to exactly align the unlensed galaxy axes with the coordinates such that $\gp$ acts along the galaxy axes, and $\gx$ acts at $45^\circ$. With the exception of mock data intentionally created with a nonzero position angle, we align our observed galaxies with the coordinate axes to best preserve this same relation. It should be noted, however, that alignment in the observed frame does not exactly correspond to alignment in the unlensed galaxy frame.}} of $\gx\approx 0.08$ was found across a sample of 358 halos from Illustris TNG simulations \citep{nelson2019illustristng}. This uncertainty represents an improvement over traditional weak lensing methods, but it remains larger than the typical available signal, underscoring the need for further refinement.

These parametric modeling techniques are reliant upon the form of the velocity profile chosen. True velocity fields do not often match such an idealized model, \rev{and one of the primary findings of DW23 was that galaxies deviate significantly from intrinsic circularity. This has an immediate consequence for the inference of shear along the galaxy axes ($\gp$ if the unlensed galaxy axes are aligned with the coordinate axes), which typically relies on an assumption of circularity when viewed face-on to determine the intrinsic axis ratio of the galaxy via the Tully--Fisher relation \citep{huff2013cosmic}. In contrast, shear applied at $45^\circ$ to the galaxy axes ($\gx$ in galaxy-aligned coordinates) induces a symmetry-breaking distortion in the velocity field that is directly observable and does not depend on any assumptions about the unlensed axis ratio.} While KL offers potential improvement over traditional weak lensing techniques on a per-source basis, an ensemble of galaxies may still be required to overcome the variability in galaxy kinematic morphology, or kinematic shape noise \citep{gurri_2020, gurri_2021}. 

To increase precision without reliance on a parametric model, \citet[][dBD15]{de_burgh-day_2015} proposed a model-independent approach, using only the symmetry properties of a galaxy to constrain the cross-component of shear. \rev{This reduces the number of fitting parameters required from $\sim 11$ for a full parametric disk model to just five (centroid position and velocity, position angle, and $\gx$)}. Later, it was shown by \citet[][DiG21]{digiorgio_novel_2021} that explicitly including the offset in the position angle between the kinematic and photometric axes, in conjunction with a kinematic model, reduces statistical uncertainty in the inferred shear parameter by a factor of 2--6. We seek to refine and extend these works, \rev{focusing exclusively on the cross-component of shear and adopting a fully model-independent framework.}

\rev{Our method, \textbf{MIRoRS} (Model-Independent Restoration of Reflection Symmetries),} takes inspiration from dBD15, but has some fundamental differences. Starting from an observed velocity field, we transform the field back to the source plane through translation, rotation, and lensing shear. When transformed with true parameter values, the symmetry (antisymmetry) under reflections across the major (minor) axis will be restored. Performing these reflections and then transforming back to the detector frame gives two new velocity fields, one for each symmetry transformation. These fields are then combined into a single nonparametric model and compared to the original data to compute a residual and a likelihood to be optimized. We also incorporate some of the methods of DiG21. In our likelihood function, we include a term from the difference between the observed photometric and modeled kinematic axes. 

We first characterize our approach using idealized mock data to explore its response and limitations. After the initial testing, we applied our model to a more realistic dataset, comprising 358 halos from the Illustris TNG simulations. This is the same dataset used by DW23, allowing for a direct comparison between our method and their parametric approach. Finally, we demonstrate an objective method for discarding outliers whose velocity fields do not follow the idealized assumptions. We utilized a Moran's $I$ test for spatial autocorrelation to search for patterns in the residuals indicating unmodeled physical effects.

The paper is structured as follows: in \S \ref{sec:Weak Lensing} we will discuss weak lensing and its effects on observables; \S \ref{sec:Methods} will outline our approach; and \S \ref{sec:Results} will cover results and discussion.

\section{Weak Lensing}\label{sec:Weak Lensing}

The theory of gravitational lensing in general, and weak lensing in particular, is well-established in the literature \citep{bartelmann_2001, bartelmann_2017, umetsu2020cluster}. Light from a background source, following a geodesic, will be deflected as it travels through the potential well of a massive body. This deflection will produce a distorted image, encoding information from the intervening mass distribution. For the purpose of this paper, we will limit our discussion to the weak lensing regime of small deflections. We will also assume that the characteristic size of the lensing mass distribution is small compared to the distances along the geodesic and adopt a thin-lens approximation of the lens.

Under these assumptions, the transformation between the source and the observed image can be expressed as
\begin{equation}\label{eq:A_matrix}
\mathcal{A} = \begin{pmatrix}1-\kappa-\gp  &- \gx \\ -\gx & 1-\kappa+\gp\end{pmatrix},
\end{equation}
where $\gp$ and $\gx$ represent the two components of the shear vector $\gamma$ and $\kappa$ gives the convergence. In our convention, the shear vector is oriented perpendicular to the line connecting the source and lens, as projected onto the plane of the sky. This transformation actually represents the inverse of the shearing operation, taking an image in the detector plane back to the source plane. Lensing of a source to the observed image is done through the inverse operation $A^{-1}$, 
\begin{equation}\label{eq:A_inv_matrix}
\mathcal{A}^{-1} = \mu\begin{pmatrix}1-\kappa+\gp  & \gx \\ \gx & 1-\kappa-\gp\end{pmatrix},
\end{equation}
where the prefactor $\mu = ((1-\kappa)^2 -\lvert\gamma\rvert^2)^{-1}$ gives the magnification, and the magnitude of total shear is given by $\lvert\gamma\rvert^2 = \gp^2 + \gx^2$. Examining this tensor gives insight into the effect of each component. Convergence produces an isotropic focusing of the image while both components of shear produce anisotropic effects. The first component, $\gp$, stretches and compresses the image along the coordinate axes. This produces an elliptical image of an otherwise circular source, but is indistinguishable from the inherent ellipticity of the unlensed object. The second component, $\gx$, shears the object along the $y=x$ axis.

\subsection{Distortion of Observables}\label{subsec:Distortion of Observables}

\begin{figure}[t]
    \centering
    \includegraphics[width=.98\columnwidth]{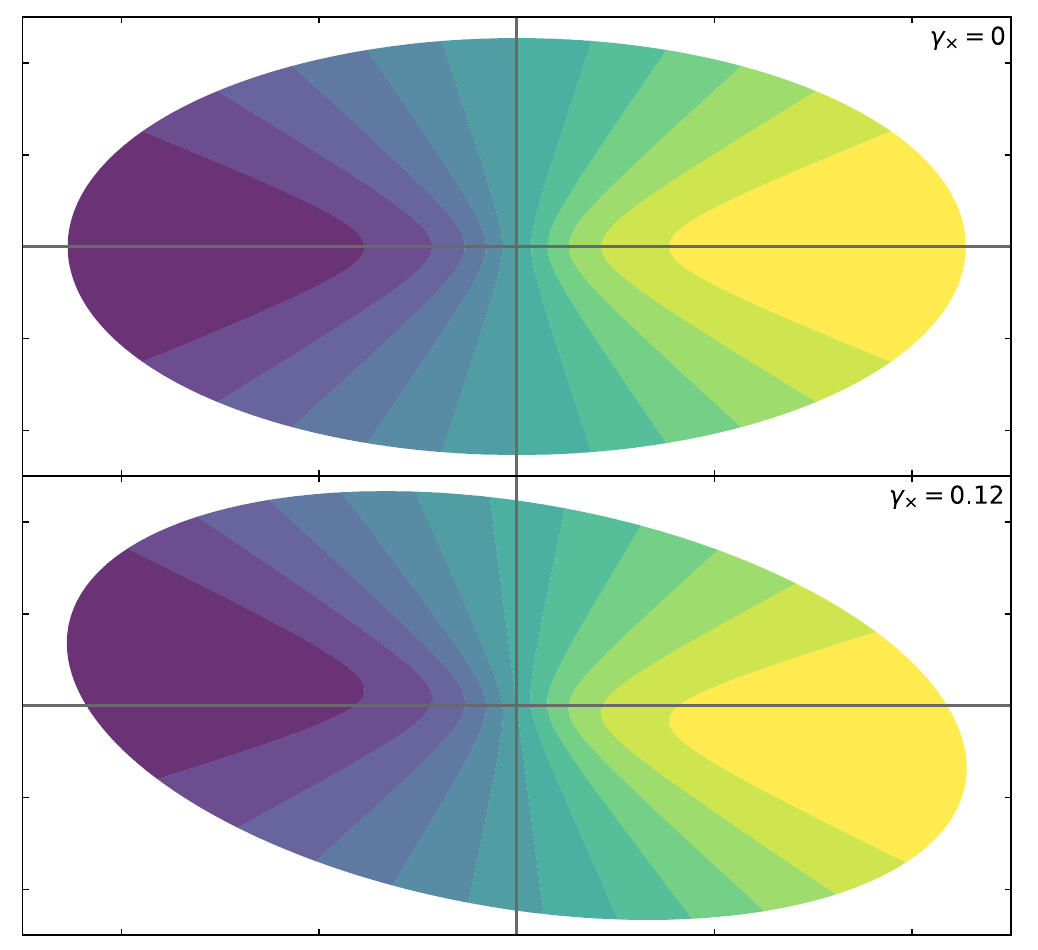}
    \caption{Velocity field of a nonsheared galaxy (top panel) and galaxy with $ \gx =0.12$ shear applied (bottom panel) from a lens along the $y=x$ line from the source. The sheared galaxy displays a notable loss of symmetry in its kinematic field as well as an apparent rotation in its morphology.}
    \label{fig:symmetry}
\end{figure}

At the heart of the \rev{MIRoRS} method lies the fact that under a weak lensing shear transformation, a disk galaxy will lose its inherent symmetries. As can be seen in \cref{fig:symmetry}, the nonsheared field exhibits a symmetry in its velocity field about the major axis and an antisymmetry across the minor axis. After applying the cross-component of shear ($\gx =0.12$ in this case), these symmetries are both lost. We use this loss of symmetry to fit for the unknown shear parameter directly by finding the transformation of the sheared field that restores the symmetry.

\begin{figure}[t]
    \centering
    \includegraphics[width=.98\columnwidth]{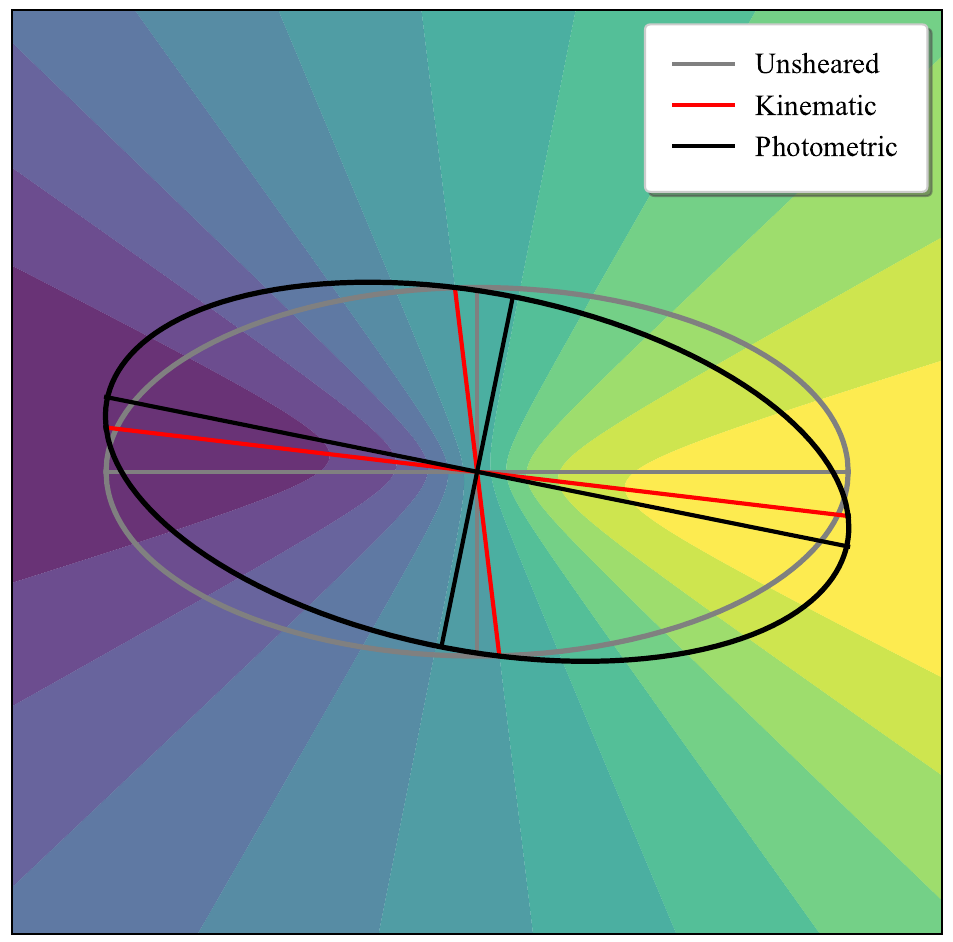}
    \caption{Pre- and post-lensing kinematic and photometric axes corresponding to an elliptical isophote overlaid on the sheared velocity field. Prior to lensing, both the kinematic and photometric axes are aligned (gray). After lensing, the isophote will remain elliptical with orthogonal photometric axes (black). For small values of $\gp$, the transformation of the isophote can be well modeled as an ellipse rotated by an angle $\theta$ as shown in \cref{eq:photo_axes}. The kinematic axes (red), on the other hand, do not remain orthogonal. The discrepancy between the kinematic and photometric major and minor axes is given by \cref{eq:delta_maj} and \cref{eq:delta_min} respectively.} 
    \label{fig:axes_symmetry}
\end{figure}

We can also take advantage of the fact that the kinematic and photometric axes are transformed differently, as can be seen in \cref{fig:axes_symmetry}. Following the approach of DiG21, an isophote of the unlensed galaxy can be modeled as an ellipse in the source ($x$, $y$) coordinates, 
\begin{equation}\label{eq:ellipse}
    1 = q^2x^2 + y^2,
\end{equation}
where q is the ratio between minor and major axes, $q = b/a$, in the source plane. We are interested in the image produced in the detector coordinates, $(x',y')$, which can be found by applying the shear transformation, $\bm{x} = \mathcal{A}\bm{x}'$. In the weak lensing regime, we expect the shear parameters to be small, allowing us to expand and drop any quadratic terms yielding
\begin{equation}\label{eq:det_ellipse_red}
    1 \approx q^2(1-2\gp)x'^2 + (1+2\gp)y'^2 -2\gx x'y'(1+q^2).
\end{equation}

Next, we will approximate the transformation of the isophote as a rotation rather than a shear. Again, in the weak lensing regime, we expect this rotation angle to be small and thus use small angle approximations to first order,
\begin{equation}\label{eq:ellipse_rot_red}
    1 \approx q'^2 x'^2 + y'^2 -2x'y'\Delta\theta(1-q'^2).
\end{equation}
Here, the observed axis ratio, $q'$, is the same in both the source and detector frames. We denote the photometric rotation angle as $\Delta\theta$ to emphasize that this rotation is the change in photometric position angle due to lensing.

Comparing coefficients from \cref{eq:det_ellipse_red} and \cref{eq:ellipse_rot_red} provides relationships between the rotation angle, $\theta$, and the shear parameter $\gx$ as well as relationships between the axis ratio in each frame. Looking first at the $y'^2$ term,
\begin{equation}\label{eq:y_coeff}
    1 = (1+2\gp),
\end{equation}
we see that $\gp \approx 0$ for the transformation to be modeled as a rotation. For our purposes, this is achieved by orienting the the galaxy axes such that the shear vector aligns with $\gx$. In a real observation, $\gp$ will typically be small, but nonzero. We address this in \S \ref{subsubsec:gp}.

With $\gp = 0$, the $x'^2$ term,
\begin{equation}\label{eq:q prime}
    q'^2 = (1-2\gp)q^2,
\end{equation}
indicates the observed axis ratio, $q' = q$, the source axis ratio.  While derived in a different manner, this matches the ultimate findings of Eq. 18 in DiG21. Finally, the $x'y'$ term gives us
\begin{equation}\label{eq:photo_axes}
    \Delta\theta = \frac{1+q'^2}{1-q'^2}\gx.
\end{equation}
This is the change in position angle of the photometric axes due to lensing. However, the pre-lensed condition of both the axis ratio and position angle is unknown.

Turning our attention to the transformation of the kinematic axes, it is evident in \cref{fig:axes_symmetry} that the velocity field cannot be modeled as a simple rotation. If we consider a point on the major axis of the unlensed galaxy, $(x,0)$, that point will be transformed to $(x',y') = (\mu(1+\gp)x, \mu\gx x)$. The kinematic axis angle, post-lensing, is then displaced by an angle
\begin{equation}\label{eq:theta_maj}
        \Delta\phi_{\text{maj}} = \tan^{-1}\lt(\frac{\gx}{1+\gp}\rt)\approx  \gx,
\end{equation}
while the minor axis is displaced
\begin{equation}\label{eq:theta_min}
    \Delta\phi_{\text{min}} = \frac{\pi}{2} -\tan^{-1}\lt(\frac{1-\gp}{\gx}\rt) \approx -\gx.
\end{equation}

Knowing the displacement of each set of axes from the unlensed condition, in which we assume the axes are aligned, allows us to infer the shear from the difference between the photometric and kinematic axes. Defining our coordinate system such that the unlensed galaxy has its axes aligned with the coordinate axes, the change in position angles reduces to the observed position angles in the detector frame. Looking first at the major axis, the difference $\Delta_{\text{maj}} =\theta'- \phi'$ is given by

\begin{equation}\label{eq:delta_maj}
   \Delta_{\text{maj}}  = \lt( \frac{1+q^2}{1-q^2}\gx\rt)-\gx= \frac{2\gx q^2}{1-q^2}.
\end{equation}
Meanwhile, the difference between observed minor axes is 
\begin{equation}\label{eq:delta_min}
    \Delta_{\text{min}}= \frac{2\gx}{1-q^2}.
\end{equation}
Further, we can solve for $\gx$ in terms of the difference in angle,
\begin{align}
    \gamma_{\times, \text{maj}} &= \frac{(1-q^2)}{2q^2}\Delta_{\text{maj}}\label{eq:gamma_maj}\\
    \gamma_{\times\text{, min}} &= \frac{(1-q^2)}{2}\Delta_{\text{min}}\label{eq:gamma_min}.
\end{align}

We can also now revisit \cref{eq:A_matrix} and \cref{eq:A_inv_matrix}. Under the current approximations of weak lensing, and by defining our coordinate system such that $\lvert{\gamma}\rvert = \gx$, the transformation matrices are then
\begin{align}\label{eq:shear_matrices}
    \mathcal{A} &= \scalebox{0.95}{$\begin{pmatrix}1 &- \gx \\ -\gx & 1 \end{pmatrix}$}, & 
    \mathcal{A}^{-1} &= \scalebox{0.95}{$\mu \begin{pmatrix}1 &\gx \\ \gx & 1 \end{pmatrix}$}. 
\end{align}

\section{Methods}
\label{sec:Methods}
\subsection{Overview}\label{subsec:overview}
We modify and combine the approaches of dBD15 and DiG21, using both kinematic and photometric data to fit for the cross-component of shear, $\gx$.  

We first characterize our methods with tests on idealized mock data. Galaxy parameters for shear, position angle, inclination, and noise are varied to explore the performance and limitations of the approach. We then analyzed a set of 358 galaxies selected from Illustris TNG100-1 simulations \citep{nelson2019illustristng}. These halos are all at $z=0$ and have an expected shear value $\gx=0$. This data was used by DW23, allowing for a direct comparison between our methods and parametric modeling. 

Finally, the residual maps for each halo are analyzed for spatial autocorrelation using a Global Moran's $I$ test. Correlated patterns in the residuals indicate unmodeled physical effects and, as such, provide a method for screening the results.

\subsection{Direct Shear Mapping}\label{subsec:Direct Shear Mapping}
\rev{The MIRoRS method is based largely upon the direct shear mapping (DSM) technique first developed in dBD15.} By exploiting the effect of weak lensing on the intrinsic symmetry of the velocity field of a galaxy, they were able to fit for the shear parameter, $\gx$, in a completely model-independent fashion. While conceptually similar, our implementation is materially different, with an emphasis on accurately capturing the model uncertainties. 

\begin{deluxetable}{lcc}
    \caption{Fitting parameters}\label{tab:params}
    \tablehead{
\colhead{Parameter} & \colhead{Symbol} &\colhead{Unit}}
    \centering
\startdata
    \textbf{Shear} & $\gx$ & ... \\ \hline
    \textbf{Central Velocity} & $V_c$ & km s$^{-1}$ \\ \hline
    \textbf{Position Angle} & $\phi$ & deg \\ \hline
    \textbf{Centroid} & $(x_c,\,y_c)$ & pixel\\ \hline
\enddata
\end{deluxetable}

The velocity field of a rotationally supported disk galaxy is expected to have intrinsic symmetry across its major axis and antisymmetry across its minor, with the major and minor axes being orthogonal. After lensing, the axes are no longer orthogonal, but have a separation given by the difference between \cref{eq:theta_maj} and \cref{eq:theta_min},
\begin{equation}
    \theta_\perp \approx \frac{\pi}{2} \pm 2\gx. 
\end{equation}

The loss of symmetry is unique to the effects of $\gx$. Thus, we can fit for $\gx$ using \cref{eq:shear_matrices} by finding the transformation that restores this symmetry. This fit will also depend on nuisance parameters for the central velocity, $(x,y)$ coordinate of the centroid, and position angle, $\phi$, in the source plane. All relevant parameters are given in \cref{tab:params}.

The general approach is to use a simulated annealing optimization to fit for these parameters. A set of parameters is assumed and used to transform the field. More in-depth discussion on these transformations is described below. The transformed field is compared to the original field to compute a per-pixel $\chi^2$ value and likelihood.

\begin{figure*}
    \centering
    \includegraphics[width=\textwidth]{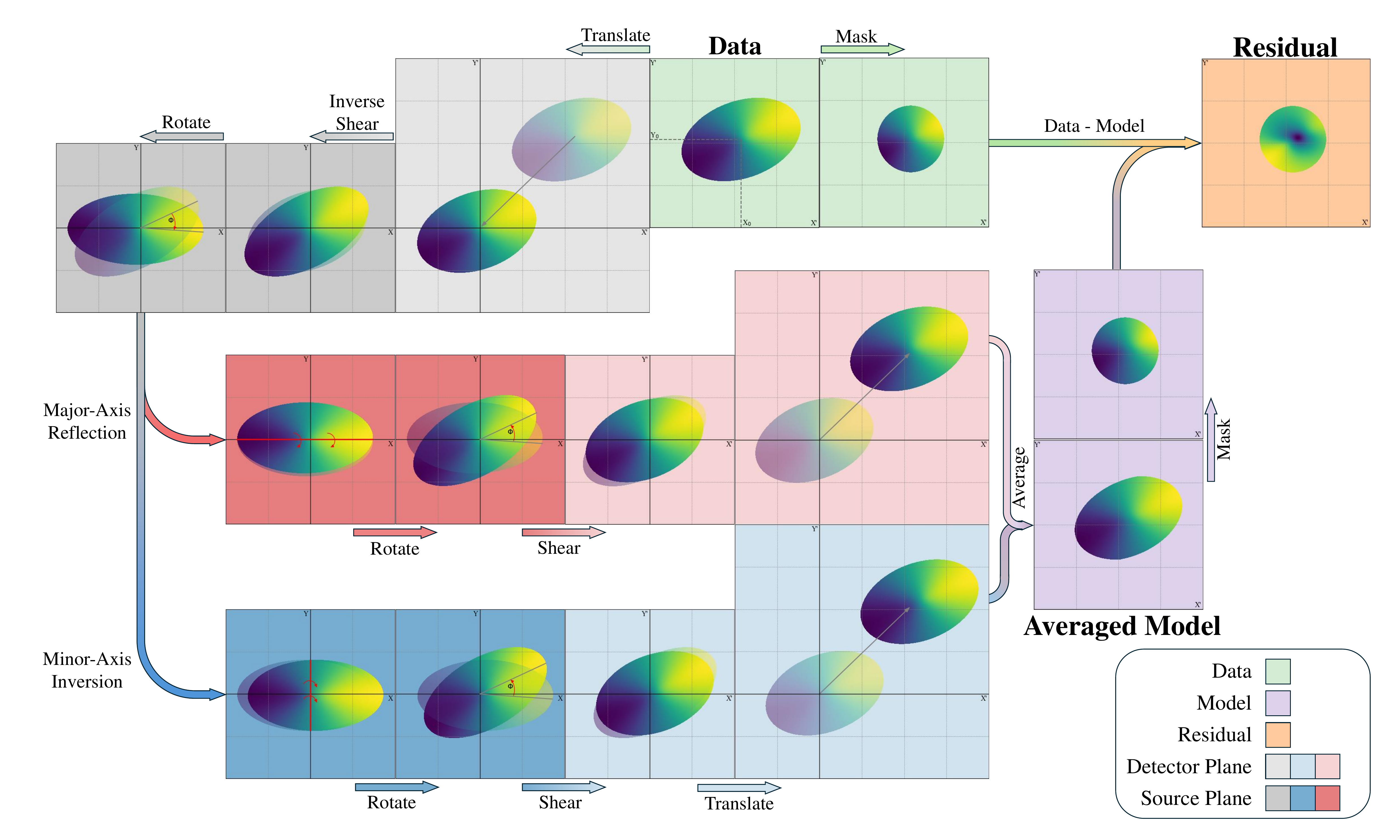}
    \caption{Illustration of the MIRoRS algorithm. Starting in the detector frame (light background), data (green) is first masked around a central pixel. To create a model, the raw data is translated to the detector origin and inverse shear is applied, bringing the image into the source plane (dark background). A clockwise rotation orients the galaxy axes with the coordinate axes. Two symmetry transformations are done independently; a reflection about the major axis (red) and an inversion about the minor (blue) using \cref{eq:v_minor_shift}. Rotation and shear are then reapplied to bring the images back to the source plane where they are translated back to the original position. The two separate images are combined into a single model (purple) and masked with the same mask as applied to the data. The difference between the model and the data produces the residual (orange). Here, the data has an applied shear, $\gx=0.15$ and position angle $\phi=30^\circ$ while the model transformations use $\gx=0.12$ and $\phi=29^\circ$ so a nonzero residual is produced. \rev{Note, the parameter values and mask sizing were chosen to provide qualitative insight and are not necessarily indicative of typical values.}}
    \label{fig:algorithm}
\end{figure*}

The transformations, shown in \cref{fig:algorithm}, start with the velocity field data and associated per-pixel uncertainty field in the observed detector frame.
\begin{enumerate}
    \item The field is first translated from its observed position to the detector origin by fitting parameters $x_c$ and $y_c$. Future operations (rotation, shear) are implemented about this origin. 
    \item Next, we apply a shear transformation to restore the unlensed fields in the source coordinates. With the correct parameters for $(x_c,y_c)$ and $\gx$, the intrinsic symmetry of the field will be restored at this point.
    \item To properly implement the symmetry transformations, the galaxy axes must be aligned with the coordinate axes. Thus, the unsheared fields are rotated by a parameter for the position angle in the source plane, $\phi$.
    \item Two independent symmetry transformations are applied to the velocity field: a reflection across the major axis and an inversion across the minor axis. To account for the antisymmetry across the minor axis, the inversion consists of a reflection followed by a sign change. 
    \item These separate transformed fields should be identical to the original field if the correct parameters are used in the transformations (assuming, of course, intrinsic symmetry). The fields are averaged into a single model. This model is then transformed back into the detector plane by applying the inverse of each operation, rotation, shear, translation. 
    \item Once back in the detector plane, the transformed model is compared to the data to compute a residual and likelihood. 
\end{enumerate}
 The actual implementation of this algorithm involves a sequence of coordinate transformations, applied to both the velocity and uncertainty fields. The likelihood produced here is combined with a term from the displacement of the photometric axes, covered in the next section.

\subsection{Photometric Axes}\label{subsec:Photometric Axes}
To constrain the shear parameter from photometric data, we need to calculate the difference between the detector-plane kinematic and photometric axes. The photometric position angle in the detector plane is directly observable. Fitting for the kinematic position angle, on the other hand, is more troublesome.  These complications are dealt with in \citet{krajnovic2006kinemetry}; however, uncertainties in kinematic position angle from this method range from  $3^\circ$ for rotational velocities exceeding $\SI{100}{\kms}$ to $17^\circ$ for rotational velocities lower than $\SI{100}{\kms}$\citep{krajnovic2011atlas3d}. 

Instead, we choose to utilize the source-plane kinematic position angle fitting parameter, $\phi$, to predict the observed photometric position angle, $\theta'$. The detector position angle can be expressed by $\phi' = \tan^{-1}\lt(\frac{y'}{x'}\rt)$, where the coordinates are expressed in the lensed frame. We can then transform these coordinates into the source plane, substituting $x=\cos\phi$ and $y=\sin\phi$, to give the detector-frame kinematic position angle as a function of fitted quantities,
\begin{equation}\label{eq:phi_prime}
    \phi'_{\text{maj}} = \tan^{-1}\lt(\frac{ \sin\phi +\gx \cos\phi }{\cos\phi + \gx\sin\phi}\rt).
\end{equation}
Combining with \cref{eq:delta_maj} and solving for the photometric major-axis position angle, we have
\begin{equation}\label{eq:phot_angle_maj}
    \theta'_{\text{maj}}= \phi'_{\text{maj}} + \frac{2q^2\gx}{(1-q^2)}.
\end{equation}
This value can be compared to the observed photometric position angle to produce a $\chi^2$ value and associated term in our likelihood function.

\subsection{Optimizer}\label{subsec:Optimizer}

We utilize the \texttt{scipy.optimize.dual\_annealing} optimizer to fit for our function parameters. The dual annealing algorithm is a global optimization method that combines the stochastic approach of simulated annealing with a local minimizer. Simulated annealing searches for the global minimum of a function by exploring the solution space through random changes. The optimizer uses a ``temperature'' to dictate the aggressiveness of its global search. Early on, at high temperature, there is considerable leniency to explore the parameter space. Over time, the temperature decreases, simulating the process of annealing in metallurgy, and the optimizer is more restricted in its search. This global search is coupled with a local search algorithm to explore the space at higher resolution. This approach allows the algorithm to escape local minima and approach the global minimum, making it particularly effective for complex, multimodal function optimization problems. In our case, we used the truncated Newton optimizer to search the local landscape. We found that overall the fit was not sensitive to initial annealing temperature or number of iterations, but did require a dedicated global search algorithm. Using a local minimizer alone would settle into a local minimum without finding the global minimum.

We also implemented Markov Chain Monte Carlo (MCMC) analysis with \texttt{emcee} \citep{foreman2013emcee} and found similar results, but at increased computation time. Because we are interested in ensemble variances rather than the variances within the fit of a particular halo, we opted for the computational efficiency of \texttt{dual\_annealing}. MCMC did help to provide valuable information about the nature of the likelihood surfaces we were exploring and would be a useful tool in analyzing individual halos and their associated uncertainties.

\subsection{Fitting Algorithm}\label{subsec:Fitting Algorithm}
\subsubsection{Kinematic Fitting}\label{subsubsec:Kinematic Fitting}
The function to be optimized is the negative log-likelihood. This likelihood is calculated from both the kinematic and photometric data with their associated uncertainties. For the kinematic calculation, we start with the observed velocity field and uncertainty map. The data is pre-processed to ensure all galaxies in the analysis have the same rotation direction, and any missing pixels have their uncertainty set to infinity.

The likelihood depends on the residual between a transformed model and the data, so, first, a data region must be defined. The transformations to create the model, particularly the shear transformation, distort the field in such a manner that pixels with no data from outside the field can be brought into the model. This necessitates using a mask region small enough that reasonable transformation parameters cannot introduce these NaN pixels. This region is selected by taking the pixels within a prescribed radius of the center of the field. Note, while the centroid location is a fitting parameter, the data region must be established prior to the fitting so that all residual calculations done by the optimizer take the difference from the same set of data.

The original velocity and error fields are then transformed to create a model via the transformations shown in \cref{fig:algorithm}. Rotations are done through the standard rotation and inverse rotation matrices,
 \begin{align}\label{eq:rot}
     \mathcal{R} &= \scalebox{0.95}{$\begin{pmatrix} \cos\phi  & -\sin\phi \\ \sin\phi & \cos\phi\end{pmatrix}$}, & \mathcal{R}^{-1} &= \scalebox{0.95}{$\begin{pmatrix} \cos\phi  & \sin\phi \\ -\sin\phi & \cos\phi\end{pmatrix}$},
 \end{align}
while the shearing operations happen via \cref{eq:shear_matrices}, stated here for completeness,
\begin{align}\label{eq:shear_mats}
    \mathcal{A} &= \scalebox{0.95}{$\begin{pmatrix}1 &- \gx \\ -\gx & 1 \end{pmatrix}$}, & 
    \mathcal{A}^{-1} &= \scalebox{0.95}{$\mu \begin{pmatrix}1 &\gx \\ \gx & 1 \end{pmatrix}$}. 
\end{align}
The two symmetry operations are similarly implemented with the matrices
\begin{align}\label{eq:reflect_mats}
    \mathcal{T}_{\text{x}} &= \begin{pmatrix}1 & 0 \\ 0 & -1 \end{pmatrix}, & 
    \mathcal{T}_{\text{y}} &=  \begin{pmatrix}-1 & 0 \\ 0 & 1 \end{pmatrix}. 
\end{align}
These matrices are combined into two overall transformation operations and applied to both the data and the error maps. Applying the transformations in a single step requires only a single interpolation, rather than an interpolation for each operation. These operations do not generally commute, so care must be taken to perform the operations in the correct sequence. First, the maps are translated to the detector origin, and inverse shear is applied to bring the observed field back to the source plane. Next, a clockwise rotation is implemented to align the major axis with the $x$-axis. The fields are then reflected across the coordinate axes, creating two sets of velocity and error maps. Each of these maps is then rotated counterclockwise and the shear reapplied. This brings the images back to the detector plane where they are translated back to the original centroid and averaged into a single model to compute a residual. Combined, these transformations are
\begin{align}\label{eq:full_transformation_mats}
    \mathcal{M}_{\text{major}} &= \mathcal{A}\mathcal{R}\mathcal{T_\text{x}}\mathcal{R}^{-1}\mathcal{A}^{-1}, \\
    \mathcal{M}_{\text{minor}} &= \mathcal{A}\mathcal{R}\mathcal{T_\text{y}}\mathcal{R}^{-1}\mathcal{A}^{-1}. \nonumber
\end{align}
While we have discussed these transformations as being done to the fields, in reality they are acting on the coordinates. The new coordinate grid to be interpolated onto, including translations, is then
\begin{align}\label{eq:translated_transformation}
    \begin{pmatrix}{x}_\text{new}\\{y}_\text{new}\end{pmatrix} =  \mathcal{M}\begin{pmatrix}{x}_\text{data}-x_c\\{y}_\text{data}-y_c\end{pmatrix} +  \begin{pmatrix}x_c\\ y_c\end{pmatrix},
\end{align}
where $\mathcal{M}$ is the transformation for either the major or minor axis reflection.

Once this transformed grid is established, cubic spline interpolation is used to produce the new velocity and error maps. The velocity field from the minor axis reflection is then scaled by the central velocity to account for the antisymmetry,
\begin{equation}\label{eq:v_minor_shift}
    V_y =  2V_c - V_\text{reflect}.
\end{equation}
The two velocity fields are then averaged to produce a model,
\begin{equation}\label{eq:avg_model}
    V_{\text{model}} = \frac{V_\text{x}+V_\text{y}}{2},
\end{equation}
while the uncertainty in the model is computed from the two uncertainty maps,
\begin{equation}
    \sigma_{\text{model}} = \frac{\sqrt{\sigma_\text{x}^2 + \sigma_\text{y}^2}}{2}.
\end{equation}
\rev{The overall per-pixel uncertainty is the quadrature sum of uncertainties in both model and data. The fields are then masked at the same position as the data, and a per-pixel $\chi^2$ and likelihood are computed. The log-likelihood function is then}
\begin{equation}
    \ln\mathcal{L} = -\frac{1}{2}\sum_i\lt(\chi_i^2 + \ln\sigma_i^2+\ln 2\pi\rt).
\end{equation}

\begin{figure}[t]
    \centering
    \includegraphics[width=\columnwidth]{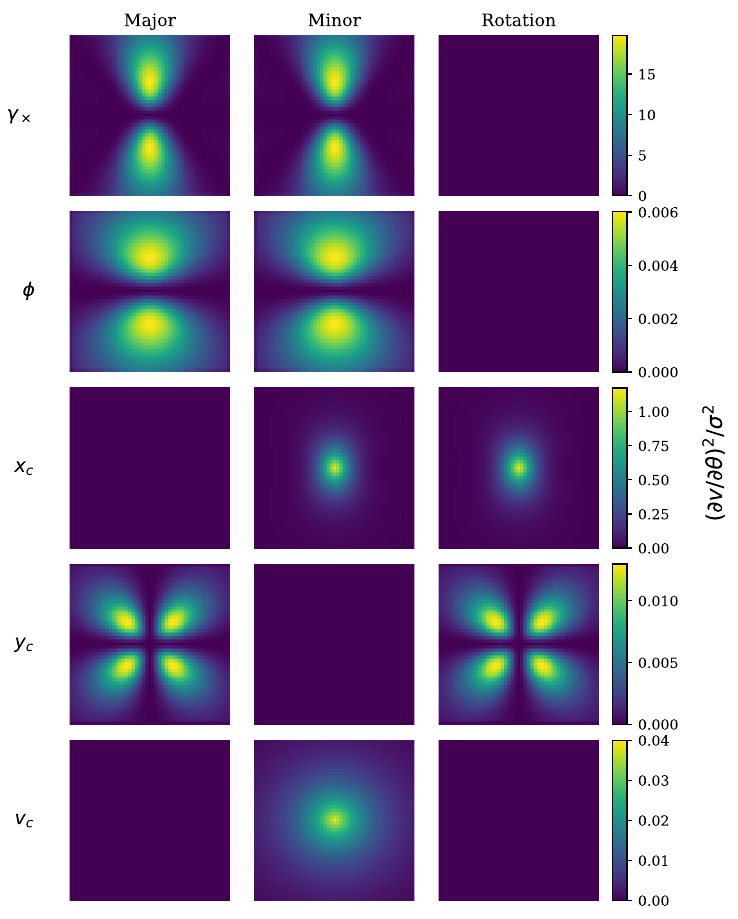}
    \caption{
\rev{Fisher information density for each fitted parameter of an ideal model ($\gamma_\times=\phi=x_c=y_c=v_c=0$), computed independently for the major-axis reflection, the minor-axis reflection, and the $180^\circ$ rotation operations. Each panel shows the spatial distribution of $(\partial v / \partial \theta)^2 / \sigma^2$ for a single parameter. The two reflection operators exhibit identical sensitivity to the shear parameter $\gx$, whereas the rotation operation provides no information on shear or position angle. The major-axis reflection constrains $y_c$ while the minor-axis reflection is sensitive to $x_c$ and $v_c$.}
}
    \label{fig:fisher}
\end{figure}

In the optimization process, each iteration will involve comparisons between models that have been created from transformations of potentially different regions of the original data with different uncertainties. To account for this variability in the model uncertainty, the normalization term, $\ln \sigma_i^2$, must be retained in the likelihood function. The third term in the expression, $\ln 2\pi$, is a constant and can be safely dropped, leaving
\begin{equation}\label{eq:log_like}
    \ln\mathcal{L} = -\frac{1}{2}\sum_i\lt(\chi_i^2 + \ln\sigma_i^2\rt).
\end{equation}

This differs from dBD15 in that no $180^\circ$ rotation operation is done, and the $\chi^2$ is computed from an averaged model instead of simply summing over total pixels. On the first issue of the rotation, we can express such a rotation by 
\begin{equation}
    \mathcal{T_\text{rot}} = \begin{pmatrix}-1 & 0 \\ 0 & -1 \end{pmatrix}.
\end{equation}
This is simply the negative of the identity matrix; the full transformation matrix is then
\begin{equation}
    \mathcal{M}_{\text{rot}} =  -\mathcal{A}\mathcal{R}\mathcal{I}\mathcal{R}^{-1}\mathcal{A}^{-1}=
 -\mathcal{I}.
\end{equation}
This holds for any shear parameter or position angle; thus, rotations are independent of shear. \rev{This is demonstrated in \cref{fig:fisher}, where we compute the Fisher information density for an ideal model with parameters $\gamma_\times=\phi=x_c=y_c=v_c=0$ and evaluate the response of the velocity field to perturbations in each parameter. The resulting maps demonstrate that the two reflections contain identical sensitivity to the shear parameter $\gx$ while rotations carry no information about the shear, position angle, or central velocity.}

\begin{figure}[t]
    \centering
    \includegraphics[width=\columnwidth]{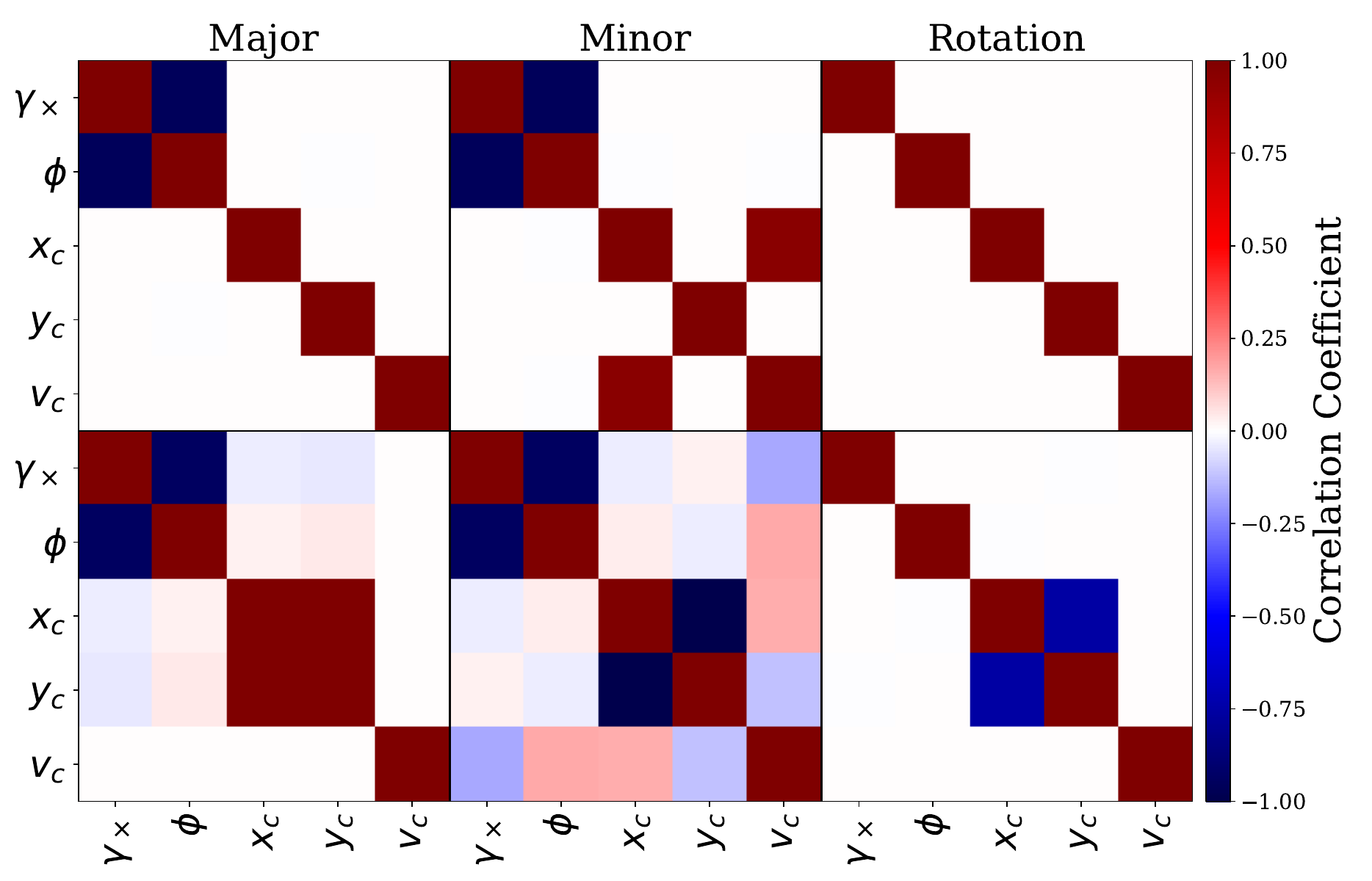}
\caption{
\rev{Correlation matrices for the fitted parameters $(\gamma_\times,\ \phi,\ x_c,\ y_c,\ v_c)$ derived from the Fisher–matrix covariance for the three symmetry operations. The top row corresponds to the ideal model used in \cref{fig:fisher}, where the expected parameter couplings appear cleanly: both reflections correlate $\gamma_\times$ with $\phi$, the minor-axis reflection couples $x_c$ and $v_c$, and the rotation operation leaves all parameters independent. The bottom row shows results for a model with nonzero parameters ($\gamma_\times=0.01$, $\phi=5^\circ$, $x_c=y_c=1$, $v_c=15\ \mathrm{km\,s^{-1}}$), which introduce additional cross-correlations, particularly for the minor-axis reflection.}
}
    \label{fig:correlation}
\end{figure}

\rev{The Fisher maps further show how the remaining parameters enter the problem: the major-axis reflection is sensitive to displacements perpendicular to that axis, constraining $y_c$, whereas the minor-axis reflection constrains $x_c$ as well as $v_c$. The two reflections show more complicated sensitivities when the parameters are not zero. This is demonstrated in the correlation matrices given in \cref{fig:correlation}, where the top row of panels shows the correlations for the ideal model used in \cref{fig:fisher} and the bottom row shows the correlations for a model created with nonzero parameters $\gamma_\times=0.01$, $\phi=5^\circ$, $x_c=y_c=1$, and $v_c=15\ \mathrm{km\,s^{-1}}$. In the top row, both reflections show a correlation between $\gx$ and $\phi$ while the rotation operation, with no constraining power on $\gx$ or $\phi$, does not show this coupling. The minor-axis operation also shows a coupling between $x_c$ and $v_c$, as expected. Both the Fisher density and correlation matrices suggest that the reflection operations carry the same information (apart from $v_c$). However, with nontrivial model parameters, it is evident that the operations have differing sensitivities and constraining power. Cross-correlations between parameters emerge differently for each operation, with the major-axis reflection remaining insensitive to $v_c$ and rotations only sensitive to centroid position. These distinct but complementary sensitivities motivate the use of the two reflections for parameter inference while omitting the rotation, which would contribute noise without adding information about $\gx$.}

On the second issue of creating an averaged model versus separately computing a $\chi^2$ for the individual fields produced by each reflection, we find that summing over the pixels of the fields separately underestimates the error due to the individual fields not being truly independent. Consider the case of the reflection across the major axis alone. The residual is the difference between data and model, but the model itself is built from the data. If we look at a simplified, two-pixel image, with one upper and one lower pixel, identified with subscripts $U$ and $L$ respectively, the $\chi^2$ value (using unit uncertainty in the data) would be
\begin{equation}
    \chi^2 = (M_U - D_U)^2 + (M_L - D_L)^2,
\end{equation}
where $M$ and $D$ refer to model and data. However, the model is just the reflection of the data such that $M_{U/L} = D_{L/U}$. With this, the double-counting is evident
\begin{align}
    \chi^2 &= (M_U - D_U)^2 + (D_U - M_U)^2\\
    &= 2(M_U - D_U)^2 = 2(M_L - D_L)^2.
\end{align}
Alternatively, rather than reflecting the data, the operation can be thought of as a folding of the data across its axis, demonstrating that the number of true comparisons is only half the pixels. The same concept holds true for the reflection across the minor axis. By summing over the entirety of pixels in each transformed field separately, the number of independent data points is artificially inflated. Instead, we choose to average over two transformations, which creates a unique comparison for each pixel in the image. 

\rev{Creating an averaged model produces the exact same result as computing separate $\chi^2$ values and manually accounting for the double-counting of pixels. We thus adopt the averaged model for simplicity while preserving the full information content of both symmetry operations.}

\subsubsection{Photometric Fitting}\label{subsubsec:Photometric Fitting}
To the likelihood function in \cref{eq:log_like}, we add a contribution from the photometric position angle. To characterize the photometric axes of a galaxy, we begin by computing the centroid of the intensity distribution. While the derivation of the angular offset in the kinematic and photometric axes does presume a colocated centroid for the two observations, we choose to calculate the photometric centroid independently rather than use the fitting parameters for the kinematic centroid, $(x_c, y_c)$. This saves computation time, calculating a single value for the photometric position angle rather than a calculation at each iteration. It also prevents the photometric observation from having an influence on the kinematic fitting parameters for $(x_c, y_c)$. The centroid coordinates $(\bar{x}, \bar{y})$ are calculated using the first-order raw moments, given by
\begin{equation}
\bar{x} = \frac{M_{10}}{M_{00}}, \quad \bar{y} = \frac{M_{01}}{M_{00}},
\end{equation}
where $M_{ij} = \sum_{x,y} x^i y^j I(x,y)$ denotes the raw moment, and $I(x,y)$ is the intensity at pixel coordinates $(x, y)$.

Next, we compute the covariance matrix,
\begin{equation}
\text{Cov}[I(x,y)] = 
\begin{pmatrix}
\frac{M_{20}}{M_{00}} - \bar{x}^2 & \frac{M_{11}}{M_{00}} - \bar{x}\bar{y} \\
\frac{M_{11}}{M_{00}} - \bar{x}\bar{y} & \frac{M_{02}}{M_{00}} - \bar{y}^2
\end{pmatrix}.
\end{equation}
Diagonalization yields eigenvalues and eigenvectors corresponding to the length and orientation of the major and minor axes. The axis ratio is inferred from the square root of the eigenvalues' ratio.

With the axis ratio, we can calculate a model value for the photometric position angle from \cref{eq:phot_angle_maj}. The axis ratio measurement introduces uncertainty in the model value, which must be accounted for. The uncertainty in the model term is calculated through standard propagation of errors
\begin{equation}\label{eq:error_prop}
    \sigma_{\theta'\text{,model}} = \sqrt{\lt(\frac{4 q \gx}{(1-q^2)^2}\sigma_q\rt)^2}.
\end{equation}
\rev{Combining this term in quadrature with the uncertainty in the observed position angle gives the total error in the photometric angle.} The uncertainties in observed position angle and axis ratio are determined directly from the distribution of position angles and axis ratios in our data sample, as shown in Section \ref{subsec:Axis Errors}. \rev{This uncertainty effectively weights the extent to which the position angle contributes to the overall fit. It was shown in DiG21 that as this uncertainty increases, the scatter in the fitted shear trends to the results from a kinematic-only fit.}
A $\chi^2$ value is then calculated
\begin{equation}
    \chi^2_{\theta'} = \frac{\lt({\theta'}_{\text{model}} -{\theta'}_{\text{obs}}\rt)^2}{\sigma_{\theta'}^2},
\end{equation}
and added to complete the likelihood function
\begin{equation}\label{eq:log_like_tot}
    \ln \mathcal{L} = -\frac{1}{2}\lt(\sum_i\lt(\chi_i^2 + \ln\sigma_i^2\rt) + \chi^2_{\theta'} + \ln\sigma_{\theta'}^2\rt).
\end{equation}
Again, the normalization term, $\ln\sigma_{\theta'}^2$, must be retained to account for changes in model uncertainty.

\subsection{Data Selection}\label{subsec:Data Selection}
\subsubsection{Idealized Mock Data}\label{subsubsec:Mock Data}
To characterize the performance of \rev{the MIRoRS method}, we first create a set of idealized mock data for both the velocity and intensity fields. The intensity field is described by an exponentially decaying profile,
\begin{equation}\label{eq:intensity}
I(r) = I_0\exp\lt(\frac{-2.99r}{r_{80}}\rt)\sec(i),
\end{equation}
where $r$ is the radial coordinate in the source plane, $r_{80}$ is the radius enclosing 80\% of the light, $i$ is the inclination angle, and $I_0$ gives the central intensity. The factor of $2.99$ is required to enforce the $r_{80}$ condition. 

For the velocity field, we model an infinitely thin disk with an arctangent profile with the major axis aligned along the $x$-axis,
\begin{equation}\label{eq:atan}
V(r) = \frac{2}{\pi}V_{\text{max}}\arctan\lt(\frac{r}{r_0}\rt)\sin(i)\frac{x}{\lvert{x}\rvert} + V_c,
\end{equation}
where $r_0$ is the scale radius independent of the intensity, $V_{\text{max}}$ is the asymptotic velocity, and $V_c$ is the central velocity of the galaxy.

Gaussian noise is added to both fields. For the intensity field, the amplitude of the noise is flat across the observation. For the velocity field, the noise is assumed to go as $I^{-1/2}$,
\begin{equation}\label{eq:vel_noise}
    \sigma_V = \delta_V\sqrt{\frac{I_0}{I}},
\end{equation}
where $\delta_V$ sets the noise level at the central pixel.

Both fields are then interpolated from the source-plane coordinates onto a grid that has been transformed by a shear parameter, $\gx$, major-axis position angle, $\phi$, and inclination $i$. 

\subsubsection{Illustris TNG}\label{subsubsec:Illustris}
After characterization with idealized data, we then analyze a set of 358 subhalos from a $z=0$ snapshot of the Illustris TNG100-1 \citep{nelson2019illustristng}  magnetohydrodynamical simulations. These halos were originally analyzed with parametric methods by DW23, and we refer the reader to their methods for an in-depth discussion on data selection. The general selection criteria required all samples to have a total mass within 10\% of $\num{1.4e12} M_\odot$, with $\geq40\%$ ($\leq 30\%$) of the stellar mass to be within the disk (bulge). 

The subhalos were all oriented according to the principal axis of the moment of inertia tensor calculated from the stellar particles. The longest principal axis was aligned along the coordinate $x$-axis as the major axis, while the second-largest principal axis was aligned as the minor axis. The galaxies were also arranged to have a clockwise rotation.

\subsection{Outlier Rejection}\label{subsec:outliers}
For a well-fit model, residuals are expected to be primarily due to uncorrelated noise. Spatially correlated patterns in residuals can be the result of a poorly fit parameter or from unmodeled physical effects such as spiral arms, bars, or warps. We use a Global Moran's $I$ test for spatial autocorrelation \citep{moran1948interpretation} to quantify the extent to which these patterns are present in the residuals, identifying halos with a poor fit. Examining residuals for spatial autocorrelation is not a new technique \citep{Goodchild1986}, but has been largely limited to fields outside of astronomy. In using the Moran's $I$ test, we seek to quantify the effectiveness of our modeling, but also to explore the utility of spatial autocorrelation in characterizing residuals.

To implement the Moran's $I$ test, we first compute the residuals by subtracting the model values from the observed values. These residuals are then normalized by their combined errors, resulting in a standardized residuals image. The Moran's $I$ statistic is then calculated to measure spatial autocorrelation, using the package \texttt{PySAL.esda} \citep{pysal2007}. Specifically, we use the Queen contiguity method (named for the directions a Queen may move in chess) to create a spatial weight matrix, where each element $w_{ij}$ is 1 if locations $i$ and $j$ are neighbors (sharing a border or vertex) and 0 otherwise. The Moran's $I$ statistic is given by 
\begin{equation}\label{eq:morans}
I = \frac{N}{W} \frac{\sum_i \sum_j w_{ij} (v_i - \bar{v})(v_j - \bar{v})}{\sum_i (v_i - \bar{v})^2}
\end{equation}
where $N$ is the number of pairwise comparisons (number of pixels squared), $W$ is the sum of all weights, $v_i$ ($v_j$) is the residual velocity value at location $i$ ($j$), and $\bar{v}$ is the mean velocity residual. This formula quantifies the degree to which similar or dissimilar values cluster together in space. A Moran's $I$ value greater than zero indicates positive spatial autocorrelation (clustering), while a value less than zero indicates negative spatial autocorrelation (dispersion). A value near zero suggests a random spatial pattern. By computing the $p$-value associated with Moran's $I$, we assess the statistical significance of the observed spatial autocorrelation against a null hypothesis of complete spatial randomness. This statistical framework enables us to identify and quantify spatially correlated residuals, taking a low $p$-value as an indication of spatial correlation in the residuals and a potentially problematic fit.

\section{Results and Discussion}\label{sec:Results}
\subsection{Axis Errors}\label{subsec:Axis Errors}

Our likelihood function contains a term involving the uncertainty in the photometric position angle and the axis ratio, so we begin by quantifying those errors. We apply our axis fitting code to the full data sample and calculate uncertainties from the resulting distributions. These empirical uncertainties are then used in the likelihood function.

First, we consider the position angles. The subhalos from Illustris are aligned with the primary principal axis of their inertia tensor along the $x$-axis. If we take this as a proxy for a ``true'' position angle of $0^\circ$, the distribution of fitted position angles represents the error in the measurement. Likewise, the position angle in the idealized mock data is known, so we are able to create a distribution of the difference between the actual and fitted values of the position angle and thus infer an uncertainty.

\begin{figure}[t]
    \centering
    \includegraphics[width=\columnwidth]{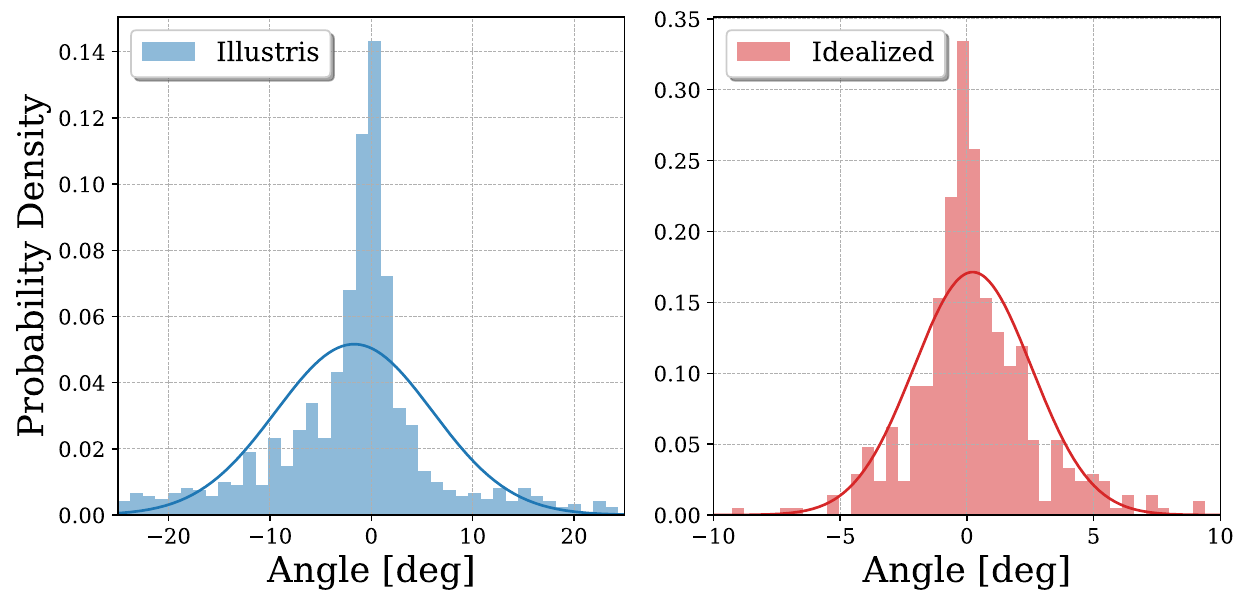}
    \caption{Distribution of errors in fitting of photometric axes. The width of the best-fit Gaussian is $7.7^\circ$ for the Illustris sample and $2.3^\circ$ for the idealized data. Both results are more sharply peaked than indicated by the Gaussian fit. } 
    \label{fig:axes_errors}
\end{figure}

We fit the resulting distributions, shown in \cref{fig:axes_errors}, with a Gaussian. Across the entire Illustris dataset, the best-fit Gaussian gives an error of $7.7^\circ$. The distribution suggests this is a conservative estimate of error, with a sharper peak in the histogram than the Gaussian fit. The approach by DiG21 assumes an error of $6^\circ$ and we find that to be in line with our results. 

We also analyzed our idealized mock data set, finding an error of $2.3^\circ$, in line with the findings of \citet{haussler2007gems} for idealized galaxies. Given the idealized nature of this data, we would expect that features like bars, warps, and spirals would produce a larger uncertainty in a real observation. However, a good fit to the ideal data provides validation of the methods used for finding the position angle. The difference between the results from Illustris and idealized data gives an indication of the impact of nonideal features, such as bars or warps, on the observed position angle.

\begin{figure}[t]
    \centering
    \includegraphics[width=\columnwidth]{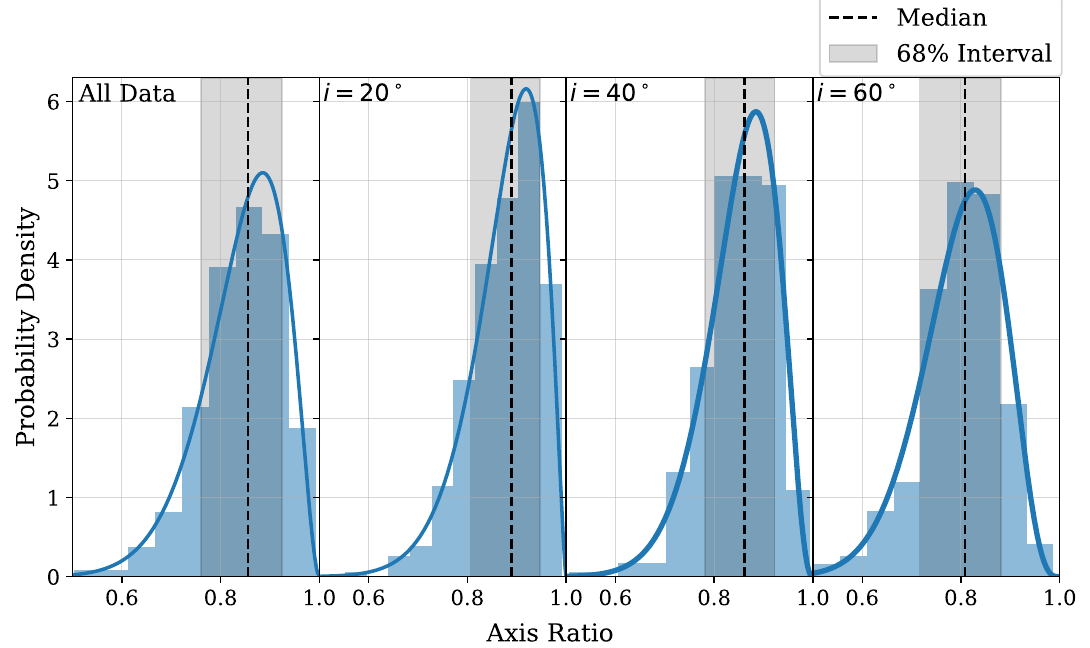}
    \caption{Axis ratios calculated from Illustris. An overall uncertainty of 0.087 was found, with similar values across inclinations of $20^\circ$, $40^\circ$, and $60^\circ$. \rev{The dashed vertical line marks the median axis ratio, and the shaded region shows the central $68\%$ interval.}}
    \label{fig:axes_ratios}
\end{figure}

Turning our attention to the errors in the axis ratio, shown in \cref{fig:axes_ratios}, we find the Illustris sample is well fit with a beta distribution across all inclination angles. An overall uncertainty of 0.087 was found over the entirety of the dataset. That value is improved to between $\sigma_q = 0.078$ and $\sigma_q = 0.085$ if we consider each inclination individually. However, that information would not be available in a real observation, so we opt to use the larger value in our calculations. Note, the axis ratio does not match what we would expect considering an inclined thin disk. Instead, all axis ratios trend greater than a theoretical value, with discrepancy increasing with inclination. We believe this to be an effect due to limitations in the ideal thin disk calculation, rather than an error in the fitting. The axis ratio in the calculation of \cref{eq:log_like_tot} depends on the true observed ratio, not on a thin disk approximation. 

Ultimately, we use an uncertainty of $6^\circ$ in the photometric position angle and $0.087$ on the axis ratio.

\begin{deluxetable}{lclr}[b]
\tablecaption{Mock Data Parameters \label{tab:mock_params}}
\tablehead{
\colhead{Parameter} & \colhead{Units} & \colhead{Range} & \colhead{Fiducial Value}
}
\startdata
$\gx$ & ... & [-0.2, 0.2] & 0 \\
$\phi$ & deg & [0, 75] & 0 \\
$i$ & deg & [10, 80] & 40 \\
$\delta_V$ & km s$^{-1}$ & [1, 40]  & 10 \\
\enddata
\end{deluxetable}

With the uncertainty in the photometric axes established, the model is then characterized on an idealized mock data set, described in Section \ref{subsubsec:Mock Data}. A set of fiducial values of $\gx = 0$, $\phi=0^\circ$, $i=40^\circ$, and $\delta_V = \SI{10}{\kms}$ were held constant across all runs, with the exception of the parameter being studied, which was varied across the range given in \cref{tab:mock_params}. A total of 50 realizations of the data are used for each parameter configuration. 

\subsection{Idealized Mock Data}\label{subsec:Idealized Mock Data}

\begin{figure*}[t]
    \centering
    \includegraphics[width=\textwidth]{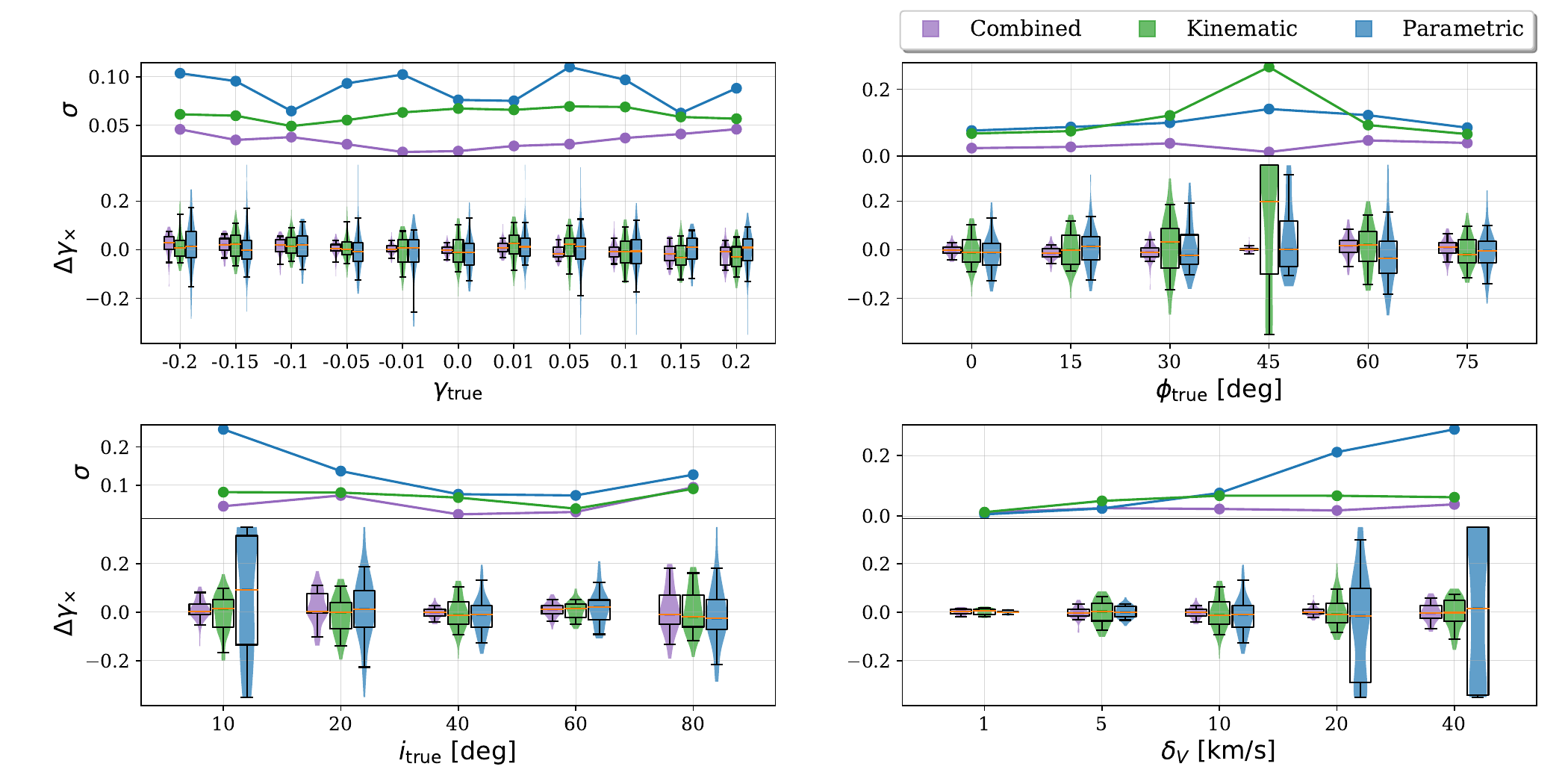}
    \caption{Characterization of the algorithm across the mock data set. A fiducial set of parameters: $\gx = 0$, $\phi=0^\circ$, $i=40^\circ$, and $\delta_V = \SI{10}{\kms}$, was used while the individual parameters were varied. \rev{In all but the top-left plot, the distribution of $\Delta \gx \equiv \gx^\mathrm{fit} - \gx^\mathrm{true}$ is equivalent to the distribution of fitted $\gx$ values for the fiducial case $\gx^\mathrm{true}=0$.} Across each trial is 50 realizations of the same configuration. In the violin plots, orange lines mark the median of the data while the boxes give the interquartile range, and whiskers extend to 5\ts{th} and 95\ts{th} percentiles. The combined model is given in purple, kinematic-only in green, and a parametric kinematic model in blue.}
    \label{fig:parameters}
\end{figure*}

We compared three approaches to fitting the data: (1) a nonparametric model using only kinematic data, (2) a nonparametric model using both kinematic and photometric data, and (3) a simple parametric model using only kinematic data based on \cref{eq:atan}, henceforth referred to as kinematic, combined, and parametric, respectively. \rev{To quantify the performance of each, we examine the distribution of $\Delta \gx \equiv \gx^\mathrm{fit} - \gx^\mathrm{true}$, which reduces to the distribution of $\gx^\mathrm{fit}$ for $\gx^\mathrm{true}=0$.} The results are shown in \cref{fig:parameters}.

While varying the value of $\gx$ from $-0.2$ to $0.2$, we find the kinematic approach generally outperforms the parametric model, and the combined model improves on the kinematic only. \rev{At larger values of shear, there is a ${\sim}{3}\sigma$ underfitting bias in the combined model.} This is a possible indication on the limits of the approximations used in deriving the theoretical difference between the kinematic and photometric position angles, particularly in the small angle approximation used in \cref{eq:ellipse_rot_red}. \rev{These systematic offsets could, in principle, be characterized with additional mock tests and used to define an empirical calibration relation for correcting the fitted shear values in future applications.}

The effects of inclination are most dramatic for the parametric model, with minimal ability to fit at the lowest inclination angle. \rev{The distribution of $\gx$ shows the parametric model running into the bounds on the parameter at $i=10^\circ$.} The kinematic model performs better than the parametric at low inclination and improves with inclination up to $60^\circ$. The combined model performs well over a wide range of inclinations, with the smallest error at mid-inclinations from $40^\circ$--$60^\circ$. Increasing to the highest inclination tested, $80^\circ$, all models have greater uncertainty than at mid-inclinations. Both nonparametric approaches give similar results at higher inclination.

The information available from the kinematic and photometric data varies with inclination. The information available in the photometric data is greatest at low inclinations. The apparent axis ratio increases with decreasing inclination, causing \cref{eq:delta_maj} to diverge as $q$ approaches unity and trend to zero as the galaxy approaches edge-on. The velocity field, however, is featureless at low inclinations. The line-of-sight velocity is a function of $\sin i$, with the differential velocity increasing with inclination. As the galaxy approaches edge-on, the minor axis vanishes in an infinitely thin disk model, causing the increase in uncertainty we see at $i=80^\circ$.

\rev{In the top-right panel of \cref{fig:parameters}, we explore the response of each model to a change in the \textit{intrinsic} (source-plane) position angle of the galaxy. The fiducial value of shear is $\lvert\gamma\rvert=\gx=0$, so all of the effects are due to behavior of the fitting algorithm in response to position angle, and not mixing of shear components. To explore this behavior, we do not make any attempt to align the galaxy axes with the coordinate axes. The fit for \gxx shows dramatic effects as the position angle approaches $45^\circ$. The kinematic-only model is completely insensitive to $\gx$ at $\phi = 45^\circ$. The shear transformation back to the source plane happens before the position-angle rotation and, with the galaxy at $45^\circ$, $\gx$ (applied with reference to the coordinate axes as opposed to the galaxy axes) will behave as $\gp$ in the galaxy frame and only produce stretching and compressing along the galaxy axes and have no effect on galaxy symmetry. This can be seen explicitly in \cref{eq:full_transformation_mats}. For the case of $\phi = 45^\circ$, 
\begin{align}\label{eq:transformations_45deg}
    \mathcal{M}_{\text{major}} &= \begin{pmatrix}0 & 1 \\ 1 & 0 \end{pmatrix}, & 
    \mathcal{M}_{\text{minor}} &=  \begin{pmatrix}0 & -1 \\ -1 & 0 \end{pmatrix},
\end{align}
completely independent of $\gx$.}

The combined model, on the other hand, seems to strongly outperform all other models at that position angle. This is \textit{not} the result of a good fit, but rather is due to the fitting process giving $\gx = 0$ for any value of true shear. If we revisit \cref{eq:phi_prime}, we can see that at $\phi = 45^\circ$, the detector-frame kinematic position angle will also be $45^\circ$. This will push $\gx$ to zero in \cref{eq:theta_maj}, regardless of true shear. This is the greatest difference in the two different nonparametric approaches, with both approaches failing for different reasons. \rev{However, in practice, the fitting can be done with the observed kinematic major axis aligned with the coordinate axis prior to fitting to minimize these effects.} 

Finally, we look at the effect of noise added to the kinematic data. Gaussian noise is added at 1, 5, 10, 20, and \SI{40}{\kms} on the central pixel and scales with the inverse of the square root of the local intensity, given by \cref{eq:vel_noise}. All models show trends of a better fit at low noise and poor fit at high noise. The combined approach increasingly outperforms the other methods as noise increases. The quality of the parametric fit at extremely low noise is probably an artifact of the mock data and parametric model using the same functional form and would not translate into performance in a real observation. \rev{At \SI{40}{\kms}, the distribution of $\gx$ shows the parametric model is running into the bounding limits and is almost completely unable to fit for shear.}

\subsubsection{Effect of \texorpdfstring{\gpp}{gamma+}}\label{subsubsec:gp}
\begin{figure*}[t]
    \centering
    \includegraphics[width=\textwidth]{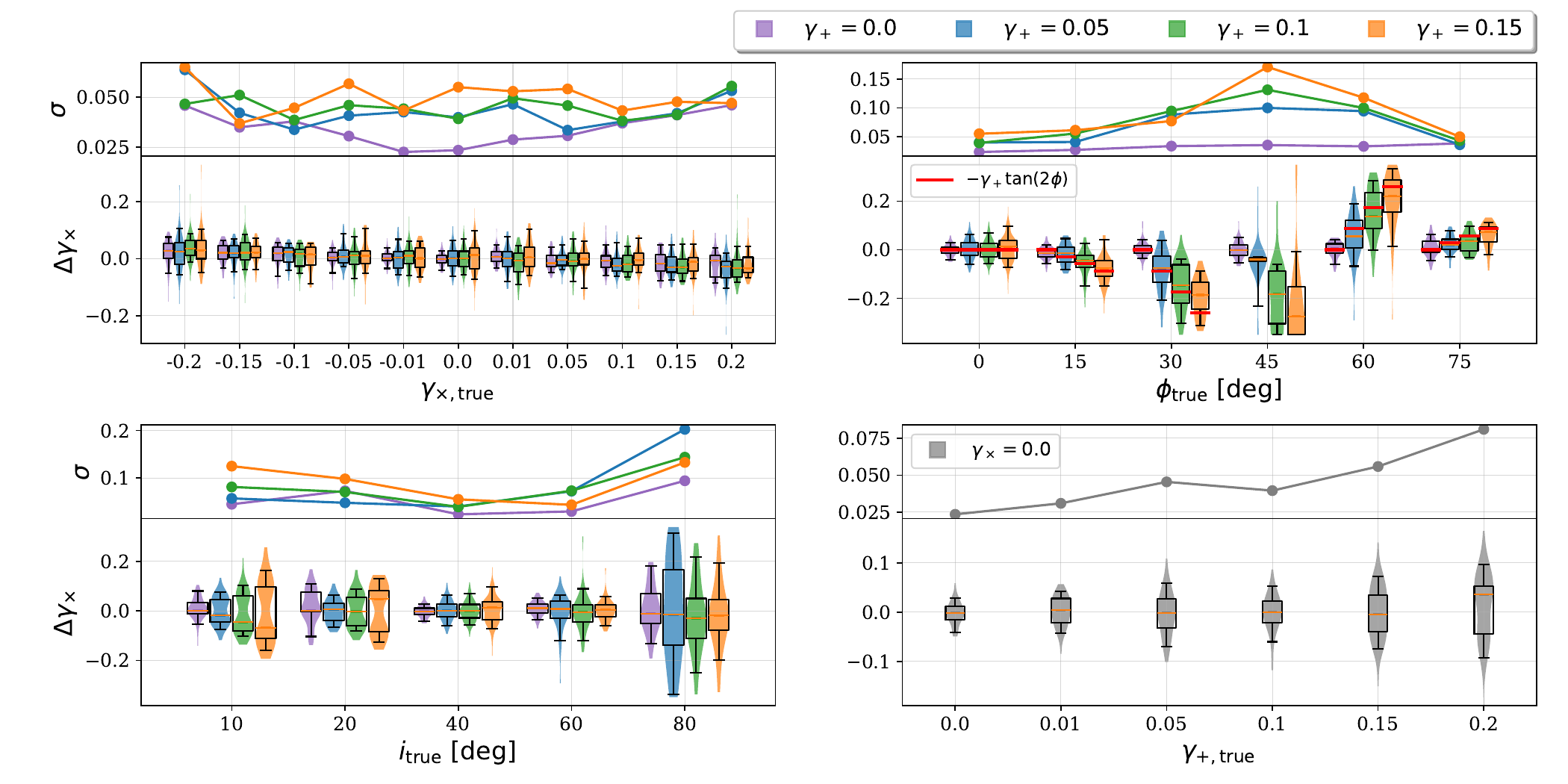}
    \caption{Impact of $\gp$ on the fit for $\gx$ using the combined model on idealized mock data. The same fiducial set of parameters: $\gx = 0$, $\phi=0^\circ$, $i=40^\circ$, and $\delta_V = \SI{10}{\kms}$, was used as in \cref{fig:parameters}. \rev{In the top-left panel, the total magnitude of shear depends on the values of both $\gx$ and $\gp$, with a maximum $\lvert\gamma\rvert = 0.25$. For all other panels, $\gx=0$, and thus $\lvert\gamma\rvert = \lvert\gp\rvert$. The top-right panel shows the effects of galaxy position angle, which effectively mixes $\gx^\mathrm{gal}$ and $\gp^\mathrm{gal}$ components.} While $\gx=0$ in the fixed detector coordinate system, it is nonzero in a coordinate system aligned with the galaxy axes. \rev{For nonzero values of $\gp$ and $\phi$, the actual value being fit is $\gx^\mathrm{fit} \simeq \gx^\mathrm{true} - \gp^\mathrm{true}\tan 2\phi$, which is plotted in red (with the exception of at $\phi=45^\circ$, where the term is undefined). The bottom-left panel shows the effects of inclination angle. At low inclination, \gpp causes the photometric axis to be vertical, which contributes to the observed bimodal distributions.} The bottom-right panel shows the effect of $\gp$ on the fiducial setup.}
    \label{fig:gp_mock}
\end{figure*}

While the MIRoRS algorithm focuses primarily on solving for $\gx$, real observations will contain both components of shear. We explore the impact of $\gp$ on our fit for $\gx$ using the combined model, using idealized mock data. \rev{We show the distribution of $\Delta \gx$ and scatter in the fit for $\gx$ in \cref{fig:gp_mock}. In all but the top-left panel, $\Delta \gx$ reduces to $\gx^\mathrm{fit}$.} We do not attempt to fit for $\gp$, but believe our methods could be improved by considering the parameter, which we leave for future work.

\rev{Overall, the effect of positive values of \gpp is to elongate the galaxy along the $y$-axis, the minor axis of our typical galaxy orientation. For small values of $\gp$, this manifests primarily as an error in the axis ratio used in fitting for the observed kinematic position angle. The top-left panel of \cref{fig:gp_mock} shows the effect of varied \gxx at each fixed value of $\gp$. The total magnitude of shear depends on both components with $\lvert\gamma\rvert = \sqrt{\gp^2 + \gx^2}$ and reaches a maximum value of $0.25$ in the case of $\gp=0.15$ and $\gx=\pm0.2$. The addition of \gpp increases scatter, particularly at low \gxx values. At larger values of $\gx$, the uncertainty is comparable across all tested values of $\gp$. The bias exhibited in \cref{fig:parameters} is also present. We can also see this increase in scatter in the bottom-right panel, where we look explicitly at the fiducial model while varying only $\gp$. For $\gp\leq0.15$, there is no bias in the fit for $\gx$, however at $\gp=0.2$, the fit is biased and actually bimodal.}

\rev{This bimodality is evidence of a distinct failure-mode of the MIRoRS method with nonzero $\gp$. The elongation of the unlensed minor axis by \gpp switches the photometric major and minor axes at sufficiently large values of $\gp$ or low inclination angles (bottom-left panel of \cref{fig:gp_mock}), causing the observed photometric major axis to align with the kinematic minor. This is a blatant violation of our assumption that the photometric transformation can be treated as a simple rotation. This produces an increased scatter as the observed $\pi/2$ difference between kinematic and photometric position angles causes the photometric term to bias the fit toward large values of shear via \cref{eq:gamma_maj}. The bimodality arises as the sign of the photometric position angle is ambiguous at $\pm \pi/2$. As large axis ratios and large values of $\gx$ are generally incompatible, this effect can be corrected by identifying the instances where shear has introduced the $\pi/2$ angle offset and defining an effective photometric position angle corresponding to the unlensed major axis. While this poses a problem that must be addressed to realize the full benefits of including the photometric position angle, \gpp (at $\phi=0^\circ$) does not affect the symmetry of the velocity field.}

\rev{The top-right panel of \cref{fig:gp_mock} shows how the source-plane position angle affects the fit for $\gx$. In the detector coordinates, a fiducial value of $\gx = 0$ is applied with varied values for $\gp$. As the position angle approaches $45^\circ$, $\gp$ applied in the fixed coordinates acts as $\gx$ in a coordinate system aligned with the galaxy axes. The MIRoRS algorithm, as stated in \cref{eq:full_transformation_mats}, first applies inverse shear to bring the observed galaxy back to the intrinsic source plane and then applies rotation by $\phi$ to align the galaxy axes with the coordinate axes before the reflection operations. Thus, the fitted components of shear are defined in the fixed detector frame, and $\phi$ alone is unable to compensate for changes in shear orientation.}

\rev{Ultimately, the optimizer is attempting to compensate for the symmetry-breaking shear component in the galaxy frame with \gxx applied in the detector frame. For model data created with \cref{eq:A_matrix}, the full transformation matrix to first order in shear, including rotations, is
\begin{equation}\left(
\begin{smallmatrix}
        \mathrm{c} 2\phi
        &
        \mathrm{s} 2\phi - 2\gp\mathrm{s} 2\phi + 2\gx\mathrm{c}2\phi
        \\
        \mathrm{s}2\phi + 2\gp\mathrm{s} 2\phi - 2\gx\mathrm{c} 2\phi
        &
        -\mathrm{c}2\phi
       \end{smallmatrix}\right),
    \label{eq:M_linear_final}
\end{equation}
where s and c represent sine and cosine operations. In contrast, the fitting algorithm, without access to $\gp$, is fitting with
\begin{equation}\left(
\begin{smallmatrix}
        \mathrm{c} 2\phi
        &
        \mathrm{s} 2\phi  + 2\gx\mathrm{c}2\phi
        \\
        \mathrm{s}2\phi - 2\gx\mathrm{c} 2\phi
        &
        -\mathrm{c}2\phi
       \end{smallmatrix}\right).
    \label{eq:M_linear_fit}
\end{equation}
Our optimizer effectively seeks to minimize the difference between the two transformations and, from the cross-terms, produces
\begin{equation}
    -\gp^\mathrm{true}\sin 2\phi + \gx^\mathrm{true}\cos 2\phi - \gx^\mathrm{fit}\cos 2\phi \simeq 0.
    \label{eq:DeltaM12_zero}
\end{equation}
The actual quantity being fit is then
\begin{equation}
    \gx^\mathrm{fit} \simeq \gx^\mathrm{true} - \gp^\mathrm{true}\tan 2\phi,
    \label{eq:gamma2_fit_general}
\end{equation}
where $\gx^\mathrm{true}$ and $\gp^\mathrm{true}$ are the true detector-frame values. This quantity, with $\gx^\mathrm{true}=0$, is shown in the top-right panel of \cref{fig:gp_mock} as red lines. While the MIRoRS algorithm fits for this theoretical value well, care should be taken to recover $\gx$ independent of \gpp and $\phi$.

To fit for the true cross-component of shear, the galaxy must be aligned such that $\phi=0$ (assuming \gpp is nonzero). Assuming equal magnitudes of \gpp and $\gx$, even small values of $\phi$ can bias the fit ($5^\circ$ would produce a bias of ${\approx}17\%$). However, $\phi$ is the source-plane position angle and is not directly observable. Notably, aligning the coordinates with the observed photometric axes can introduce significant bias, given by \cref{eq:photo_axes}. Aligning with the observed kinematic axes produces a much smaller bias with $\phi\approx\gx$. 

\begin{figure}[t]
    \centering
    \includegraphics[width=\columnwidth]{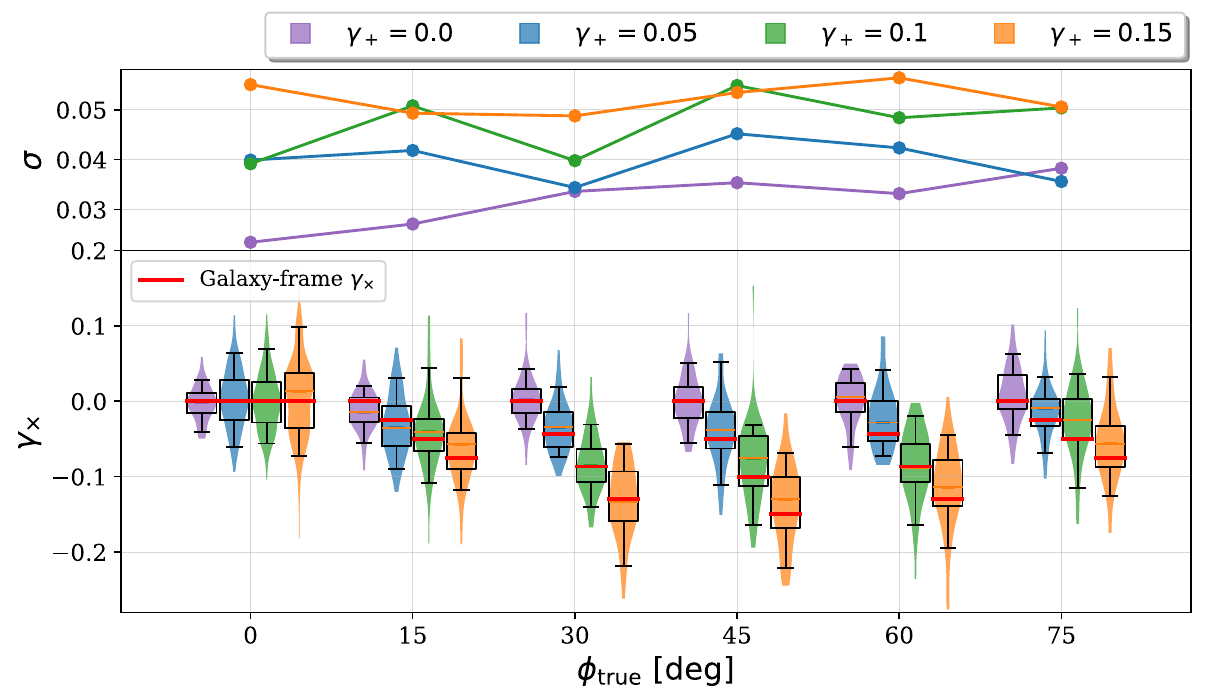}
    \caption{Distribution of $\gx$ for mock data constructed with nonzero \gpp and $\phi$ after an initial rotation to minimize $\phi$ and capture shear in the galaxy coordinate frame. The true galaxy frame \gxx is shown in red.}
    \label{fig:gp_mock_rotate}
\end{figure}

To explore these effects, we apply an initial rotation to align the detector coordinate $x$-axis with the observed kinematic axis and fit with the standard MIRoRS algorithm. We fit for the kinematic major axis by maximizing a directional gradient, but admit more sophisticated methods are likely better (see \citet{krajnovic2006kinemetry}). Adding an additional fitting parameter for the initial rotation is largely degenerate with the rotation in the source plane, complicating the fit. However, it is likely possible that an iterative process could be used that starts at the observed kinematic position angle and applies a rotation according to \cref{eq:delta_maj} until the fitted value of $\phi$ is pushed to zero, which we leave for future work. The resulting fit is shown in \cref{fig:gp_mock_rotate}. For most values of $\gp$ and $\phi$, the fitted value matches the expected value of $\gx$ in the galaxy reference frame. At the largest values of applied shear, there is an underfitting, matching what is was seen in the top-left panels of \cref{fig:parameters,fig:gp_mock}.}

Overall, when fit with the combined model, the uncertainty in $\gx$ with $\gp < 0.2$ applied remains an improvement over both the parametric and kinematic models with no $\gp$ applied. In a real observation, the shear from $\gp$ is expected to be small, but efforts to include the parameter would be warranted for a complete treatment. 

\subsubsection{Effect of Asymmetries}\label{subsubsec:asymmetries}

\begin{figure}[t]
    \centering
    \includegraphics[width=\columnwidth]{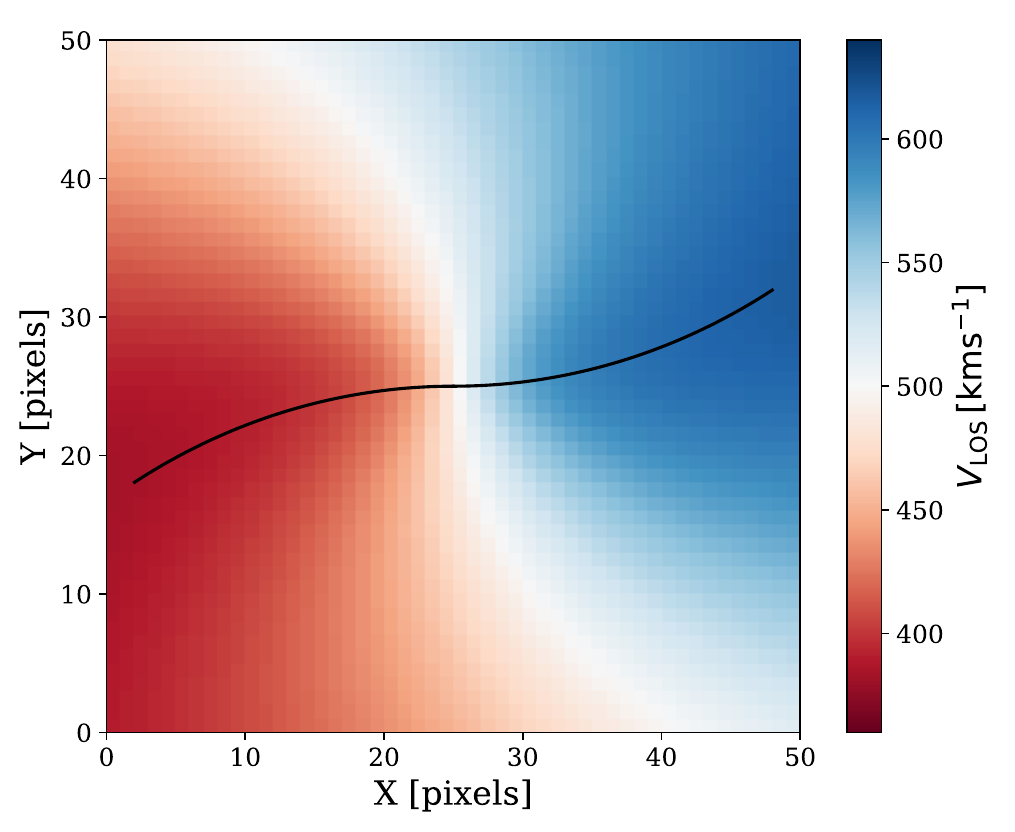}
    \caption{\rev{Example of a mock velocity field with an imposed linear drift in kinematic position angle used to test sensitivity to spatially correlated structure. The curved line marks the kinematic major axis as a function of radius for a variation of $35^\circ$ across the field.}}
    \label{fig:warped}
\end{figure}

\begin{figure}[t]
    \centering
    \includegraphics[width=\columnwidth]{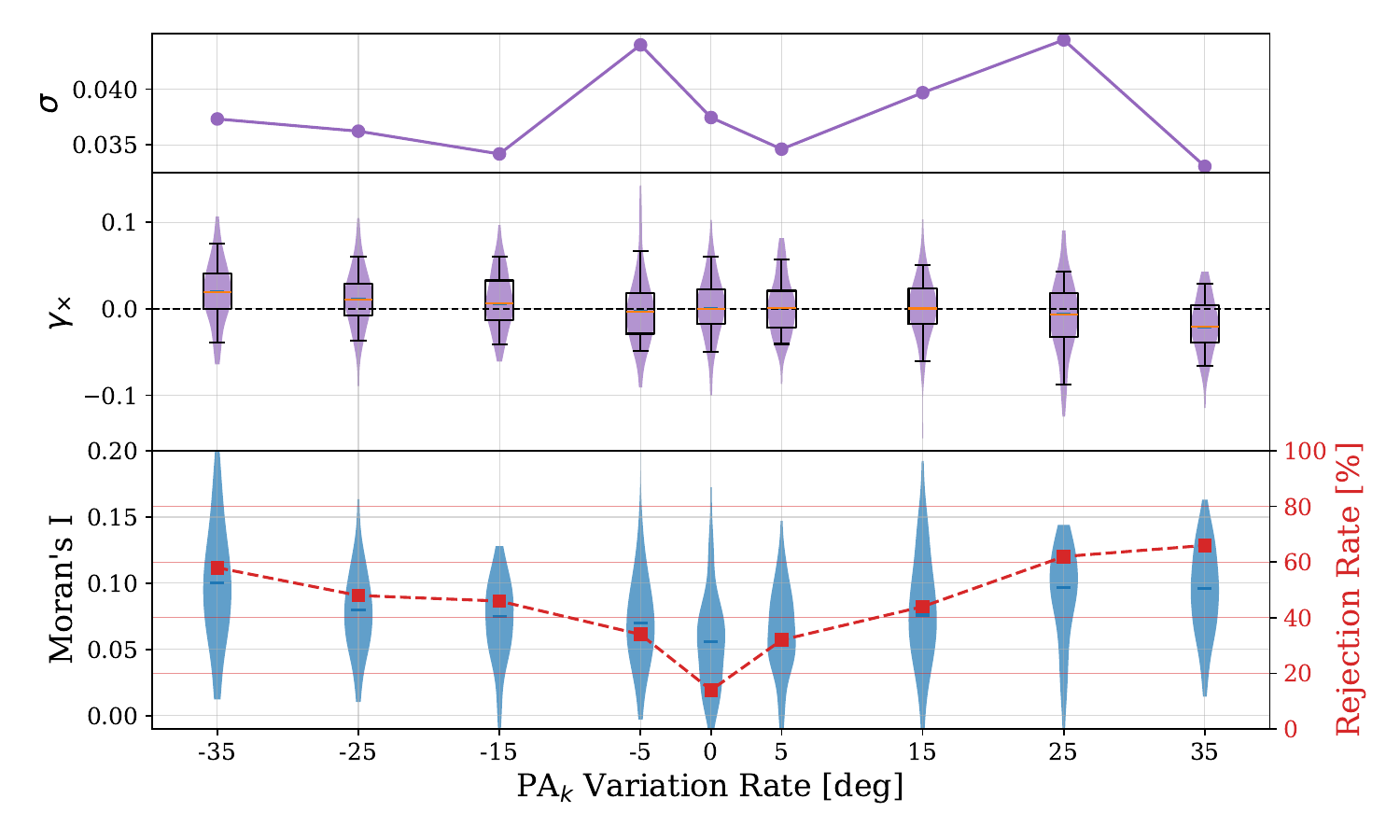}
    \caption{\rev{Characterization of the MIRoRS algorithm on mock galaxies with a linearly varying kinematic position angle. \emph{Top panel: }fitted scatter across different imposed PA-variation rates. The middle panel shows the distribution of fitted $\gx$ values and corresponding bias, which remains small but nonzero due to the coherent PA drift. The bottom panel shows Moran's $I$ (blue violins) and the fraction of fits rejected at the $p\leq10^{-3}$ level (red dashed curve).}}
    \label{fig:warped_mock}
\end{figure}

\rev{To explore the effects of spatially correlated asymmetries, we created mock velocity fields with a kinematic position angle that varies linearly with radius, $\mathrm{PA}(r)=\phi+\mathrm{PA}_k\,r$, where $\mathrm{PA}_k$ is the imposed PA-variation rate. This produces a velocity field (see \cref{fig:warped}) that violates both reflection symmetries in a coherent manner, and is the most common type of asymmetry found across a survey of 2776 MaNGA galaxies \citep{stark2018sdss}. We generated 50 realizations for variation rates of $\pm5^\circ$, $\pm15^\circ$, $\pm25^\circ$, and $\pm35^\circ$ across the field using the same fiducial parameters described in \S\ref{subsec:Idealized Mock Data}. The results shown in \cref{fig:warped_mock} exhibit scatter comparable to the symmetric case, but with a systematic bias that increases with $\mathrm{PA}_k$. The largest variation rate produced a median recovered shear of $\gx = 0.0207$, a $3.5\sigma$ deviation of the median, where $\sigma$ is the standard error of the median.

We then applied the Moran's $I$ test (bottom panel of \cref{fig:warped_mock}) and found that the statistic increases monotonically with $\lvert{\mathrm{PA}_k}\rvert$, as expected for increasingly coherent departures from symmetry. Applying a selection cut of $p \leq 0.001$ rejected $\sim 70\%$ of realizations at the highest variation rates while rejecting only $\sim 15\%$ in the fiducial case, confirming the efficacy of the Moran test in identifying asymmetric or warped fields. Quantifying these effects also opens a path toward correcting them (e.g. by explicitly estimating the degree of asymmetry with a Radon transform \citep{stark2018sdss}). In any sufficiently large ensemble, these orientation-dependent effects are expected to average away.}

\subsection{Illustris Data}\label{subsec:Illustris Data}

\begin{figure*}[t]
    \centering
    \includegraphics[width=\textwidth]{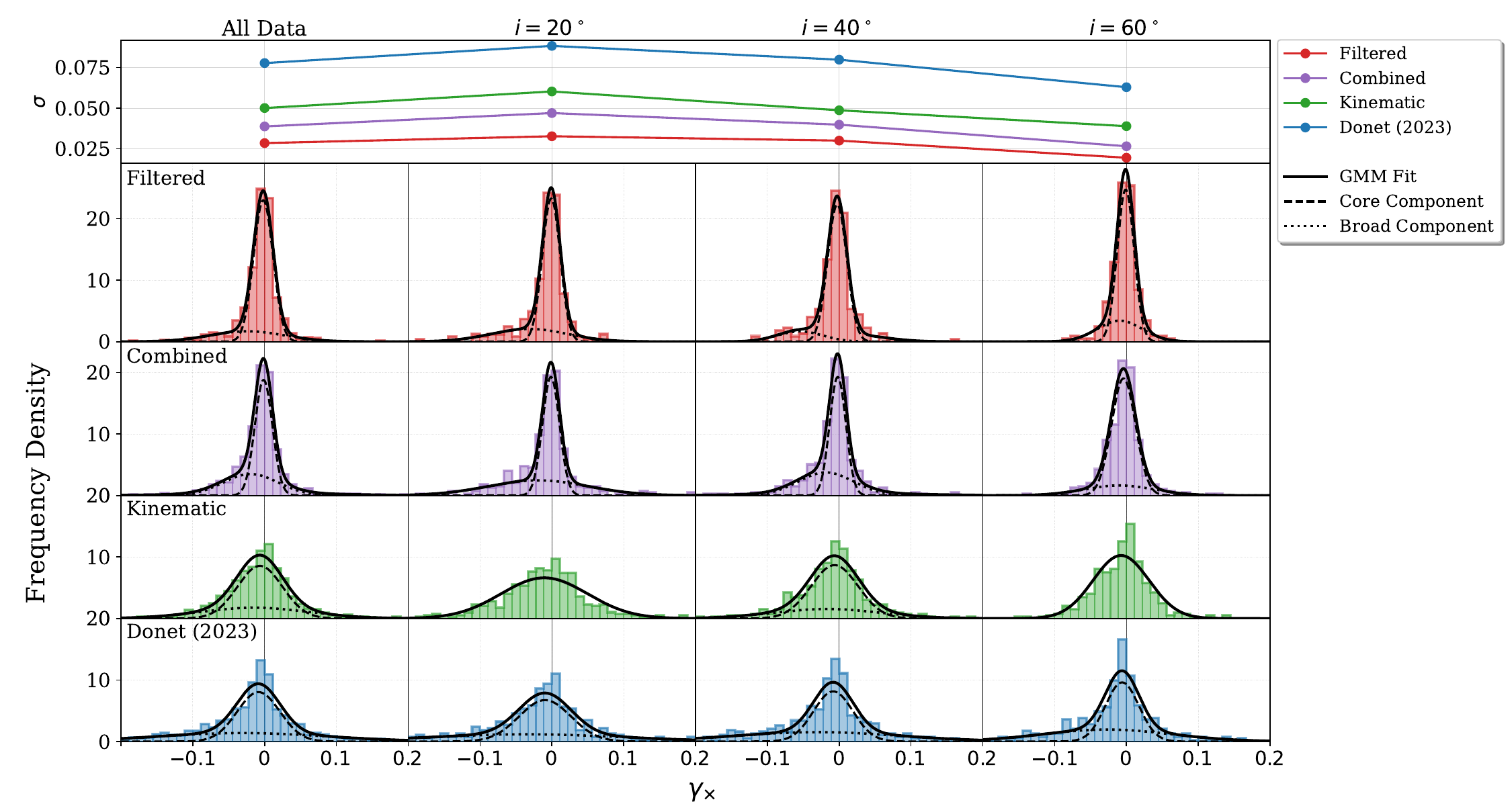}
    \caption{\rev{Comparison of fitted \gxx values for all Illustris halos at inclinations of $20^\circ$, $40^\circ$, and $60^\circ$. \emph{Top panel:} uncertainty in $\gamma_{\times}$ for each method as a function of inclination, comparing the sample filtered with Moran's $I$ test (red), the combined model (purple), the kinematic-only model (green), and the parametric model of DW23 (blue). \emph{Bottom panel:} histograms showing the distribution of fitted \gxx in each inclination bin. Each model is displayed as a row with the same color-coding as above. The preferred two-component GMM fit is marked by a solid line with the core (broad) component as a dashed (dotted) line. All histograms are plotted with the same scaling for equal comparison. The true shear for all halos is $\gamma_{\times}=0$.}
}
    \label{fig:violin}
\end{figure*}
Extending our analysis to the Illustris data, we fit the 358 halos at three inclinations each ($20^\circ$, $40^\circ$, and $60^\circ$) with both the kinematic and combined models. These halos are from a z=0 snapshot and are all expected to have $\gx=0$. Results are shown in \cref{tab:stats}, along with a comparison to analysis of the same data using a parametric model by DW23 and results after filtering with the Moran's $I$ test. \rev{We report the overall uncertainty in the distribution of fitted \gxx values, but find that the majority of the results are best fit with a two-component Gaussian mixture model (GMM) according to the Bayesian information criteria. We report the scatter in the narrow component of the GMM separately as $\sigma_{\mathrm{core}}$. We show the distribution of results along with the overall uncertainty for each case in \cref{fig:violin}.}

\begin{deluxetable}{ccccccc}[t]
\caption{Comparison of Models}\label{tab:stats}
\tablehead{
\colhead{Model} & \colhead{Inclination (deg)} & \colhead{Mean} & \colhead{Median} & \colhead{$\sigma$} & \colhead{\rev{$\sigma_{\mathrm{core}}$}}
}
\startdata
\multirow{5}{*}{Filtered} 
& All & -0.007 & -0.002 & 0.028 & 0.014\\
& $20$ & -0.009 & -0.002 & 0.033 & 0.012 \\
& $40$ & -0.007 & -0.003 & 0.030 & 0.013\\
& $60$ & -0.003 & -0.002 & 0.019 & 0.012 \\
\midrule
\multirow{5}{*}{Combined} 
& All & -0.007 & -0.003 & 0.039 & 0.016 \\
& $20$ & -0.009 & -0.003 & 0.047 & 0.012\\
& $40$ & -0.008 & -0.003 & 0.040 & 0.016\\
& $60$ & -0.005 & -0.003 & 0.027 & 0.017\\
\midrule
\multirow{5}{*}{Kinematic} 
& All & -0.009 & -0.006 & 0.050 & 0.031\\
& $20$ & -0.011 & -0.008 & 0.060 & ...\\
& $40$ & -0.010 & -0.005 & 0.049 & 0.033 \\
& $60$ & -0.007 & -0.005 & 0.039 & ...\\
\midrule
\multirow{4}{*}{Donet (2023)} 
& All & -0.017 & -0.010 & 0.078 & 0.030\\
& $20$ & -0.018 & -0.012 & 0.089 & 0.036 \\
& $40$ & -0.015 & -0.010 & 0.076 & 0.029 \\
& $60$ & -0.014 & -0.009 & 0.062 & 0.025\\
\enddata
\end{deluxetable}

DW23 analyzed the same Illustris data set, allowing for a direct comparison between methods, with their results shown in blue in \cref{fig:violin}. In that paper, they fit their distribution with a double Gaussian profile with a broad component plus a narrow core. Our method largely eliminates the broad wings seen in their distribution, such that our overall scatter is comparable to the scatter in the core of their distribution. While DW23 cites this low core uncertainty at $0.025$--$0.036$, their total uncertainty is between $0.089$ and $0.062$, depending on inclination. Across the same inclinations, we report a total uncertainty of $0.027$--$0.047$ with an overall uncertainty of $\sigma = 0.039$ across the entire sample. \rev{Our comparable core uncertainties are roughly half the overall, at $0.016$ for the combined model across all inclinations (\cref{tab:stats}).}

Comparing the two nonparametric models, shown in purple and green in \cref{fig:violin}, we see the combined model shows a distinct improvement over the kinematic model. Total uncertainty in the fit of \gxx with the kinematic model is $0.05$ considering all inclination angles and improves to $0.039$ at an inclination of $60^\circ$. The combined model has an uncertainty of $0.039$, improving to $0.027$ at $i=60^\circ$. 

This improvement is not as dramatic as the factor of 2--6 claimed in DiG21. In that study, a parametric model of the velocity field was used similar to our mock field. However, they did not create a mock photometric observation and instead fed their model the true position angle perturbed with a $6^\circ$ uncertainty. While some of the results from our fitting of the mock data are within the range of a factor of 2--6 improvement, we do not find that to be the case generally, especially not with a more realistic observation. 

We can also see a slight bias toward negative shear, with all mean and median values being negative. The standard error in the mean is $0.001$ when including all inclinations. With mean and median values of $-0.007$ and $-0.003$ for our combined model, this result is significant at 3$\sigma$--7$\sigma$ and matches the findings of DW23. As in that paper, we oriented all galaxies with a clockwise rotation ($Z$-wise orientation of spiral arms), \rev{producing a bias that matches our findings for warped data in \S\ref{subsubsec:asymmetries}}. In a real observation averaged over an ensemble of galaxies of arbitrary orientation, we expect the bias to vanish. However, the bias could be considered in individual observations.

\subsection{Moran's \texorpdfstring{$\mathrm{I}$}{I} Test}\label{subsec:Moran's I}
\begin{figure}[t]
    \centering
    \includegraphics[width=\columnwidth]{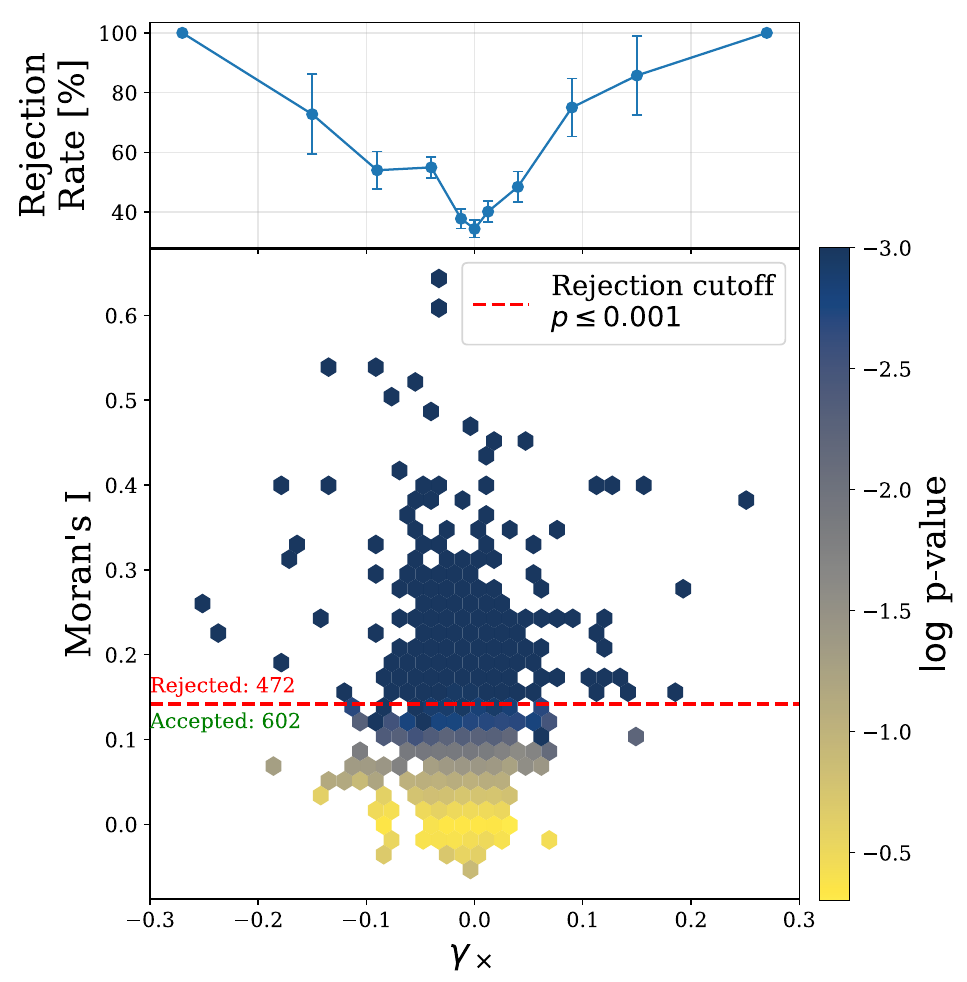}
    \caption{\rev{Moran's $I$ analysis of the Illustris sample. \emph{Top panel:} fraction of realizations rejected as a function of fitted $\gx$ when applying a significance threshold of $p \leq 0.001$. \emph{Bottom panel:} Moran's $I$ vs. fitted $\gx$ for all 1074 mock observations, with hexagonal binning colored by $\log p$–value. The red dashed line marks the rejection cutoff ($p \leq 0.001$), which removes 472 cases while retaining 602. Large values of Moran's $I$ (corresponding to low $p$–values) correlate with large inferred $|\gx|$, indicating that spatially coherent residuals are associated with biased shear estimates.}}
    \label{fig:morans}
\end{figure}

Finally, we use Moran's $I$ test to filter the Illustris results based on coherent patterns in the residuals. The \texttt{PySAL.esda} package returns both an $I$-value and a $p$-value for the null hypothesis of complete spatial randomness. In \cref{fig:morans}, we see a definite trend in the relationship between the $p$- and $I$-values, with higher $I$-values corresponding to lower $p$-values.

A good fit is one that exhibits spatial randomness, corresponding to an I-value near zero. Given the strong relationship between the p and I-values, we choose to filter the results based on $p$-value alone, rejecting results with $p \leq 0.001$. We can validate this choice by examining how the I and $p$-values relate to our fit for $\gx$, \cref{fig:morans}. There is a definitive trend for large value $p$-values and small I-values when $\gx$ is near its true value of $0$. 

Filtering the results of the Illustris data fit with the combined model, shown in red in \cref{fig:violin}, produces a noticeable improvement in the scatter. The 358 halos at three inclinations each produce a total $1074$ mock observations. Of these, $472$ were filtered out, leaving $602$ with no consideration for inclination angle. The uncertainty in the sample was reduced from $0.039$ to $0.028$ with improvements at all inclination levels, shown in \cref{tab:stats}.

\rev{The final distribution of fitted \gxx values, after filtering, is shown in red in \cref{fig:violin}. Across all inclinations, the filtered core uncertainty is $0.014$, half the total uncertainty. The negative bias in the fit remains and dominates the fit of the broad component in all but the $i=60^\circ$ case. These galaxies can be identified in \cref{fig:morans} as the cluster of galaxies with $\gx\approx-0.1$ and $I<0.1$. It is also evident that these galaxies would remain in the filtered sample for any small adjustment in the filtering threshold.}

In a real observation, the most likely cause of coherent patterns in the residual will be unmodeled physical effects. Features such as warps or bars will not display the same symmetries as an ideal disk. Filtering with Moran's $I$ test can help to identify targets that are poorly fit due to these features. We choose to simply eliminate these candidates from our analysis, but it is conceivably possible to eliminate the problematic features through creative masking to select the most symmetric regions of the galaxy.

\begin{figure*}[t]
    \centering
    \includegraphics[width=\textwidth]{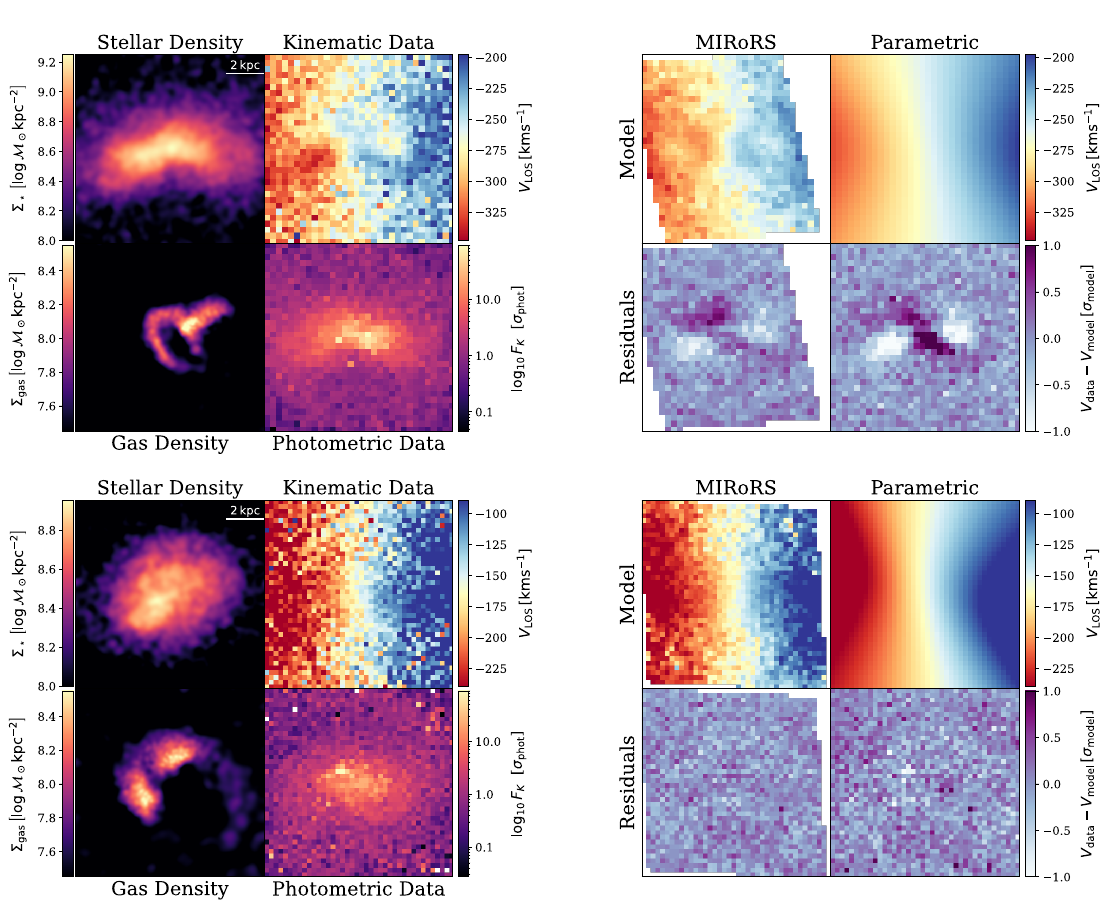}
    \caption{Visualization of Halos 491282 and 483113 and comparison of \gxx fit with the MIRoRS model and with the methods of DW23. \emph{Top panel:} results for Halo 491282. The four left panels show the stellar and gas column densities from Illustris along with the derived velocity and intensity fields. The set of four panels on the right shows the models and residuals from fitting with MIRoRS and a parametric algorithm based on DW23. The prominent substructure is unable to be modeled using symmetry operations and appears as a pattern in the resulting residual after fitting with a shear parameter of $\gx = -0.0396$ with MIRoRS and $\gx=-0.0989$ in DW23. This halo was filtered out of the MIRoRS sample with an I-value of $0.6435$ ($I=0.7004$ for the parametric model). \rev{\emph{Bottom panel:} results for Halo 483113. The Moran test did not identify any correlated residuals in the MIRoRS fit with a value of $I=0.0601$. Shear was fit moderately well with MIRoRS ($\gx=-0.0268$), but the fit from DW23 ($\gx=-0.166$) was further from the expected value of zero.} All images are 10 kpc across.} 
    \label{fig:combined_halo}
\end{figure*}

We examine the halo from our sample with the largest Moran $I$-value. Halo $491282$, shown in the top panels of \cref{fig:combined_halo}, was fit with a shear parameter of $\gx=-0.0396$ with a Moran $I$-value of $0.6435$ and $p$-value of $0.001$ at an inclination of $20^\circ$. \rev{The right set of panels gives the model and residuals for both the MIRoRS method and the parametric approach implemented in DW23.} There are noticeable unmodeled features visible in the residuals of both fits. While the kinematic and photometric data both look fairly innocuous, upon closer inspection of the stellar and gas column densities available from Illustris (far left panels), there is distinct substructure present. These features are not symmetric and, thus, we are unable to precisely fit them with our methods. \rev{The results from DW23 produced a fit further from the expected value at $\gx=-0.0989$ and display more pronounced residuals. }

\rev{In comparison, we also examined a system that had no correlated residuals identified ($I=0.0601$) when fit with MIRoRS. Halo 483113 at $i=40^\circ$ was fit with a shear parameter of $\gx=-0.0268$ with MIRoRS while DW23 found $\gx=-0.166$. We also performed the Moran's test with the residual produced by the parametric model and find the statistic larger, but not meeting our rejection threshold ($I = 0.094$, $p=0.014$). Overall, this is an excellent example of the strengths of the MIRoRS method. Using our rejection threshold, both methods would be considered a successful fit, yet the shear parameter found with the parametric method is far from the expected truth.}

\section{Summary and Conclusion}

In this paper, we have developed a model-independent technique for fitting the cross-component of shear, $\gx$, by integrating kinematic and photometric measurements. Building on the work of dBD15 and DiG21, we employed symmetry properties of a galaxy's velocity field and photometric axes to refine the shear parameter estimation.

We first characterized our approach using idealized mock data, varying parameters for shear, position angle, inclination, and noise to explore the robustness and limitations of our method. Our nonparametric kinematic model showed improved precision over traditional parametric models, particularly when combined with photometric data. The addition of applied $\gp$ increases scatter in the fit for $\gx$ with the combined model. However, for values of $\gp$ up to 0.2, the uncertainty remains an improvement on both kinematic and nonparametric approaches tested without $\gp$ applied. \rev{Further, we demonstrate the necessity to align the galaxy major axis (kinematic, or unlensed) with the detector $x$-axis. We also explored the effect of asymmetries in the velocity field with mock data created with a linearly varying kinematic position angle and find a small systematic bias, but no increase in scatter.} 

Applying our model to a realistic dataset of 358 halos from the Illustris TNG simulations, we achieved a significant reduction in the uncertainty of $\gx$, demonstrating an overall uncertainty of $0.039$ \rev{and core uncertainty of $0.016$}. These results represent a substantial improvement over a previous parametric approach by DW23, which reported an overall uncertainty of $0.078$ and core uncertainty of $0.030$ on the same dataset.

Additionally, we introduced a method for identifying and discarding outliers using Moran’s I-test for spatial autocorrelation. This technique effectively filtered out halos with coherent patterns in their residuals, reducing the overall uncertainty to $0.028$ \rev{(core $0.014$)}.

Our findings highlight the potential of integrating kinematic and photometric data in weak lensing studies, offering a more precise measurement of weak lensing shear.

\section*{Acknowledgments}
We thank Jean Donet for his advice and assistance on the project, as well as his work in supplying all of the processed Illustris data files. We would also like to thank the anonymous referee for providing constructive feedback.

\bibliography{bib}{}
\bibliographystyle{aasjournal}

\end{document}